\documentclass[longauth]{aa}
\pdfoutput=1

\usepackage{txfonts}
\usepackage{xcolor}
\usepackage{graphicx}
\usepackage{array}
\usepackage{amsmath}
\usepackage{float}
\usepackage{hyperref}
\usepackage{ulem}
\usepackage{makecell}
\usepackage{upgreek}
\usepackage{makecell}
\usepackage{lineno}
\usepackage{placeins}

\hypersetup{
    colorlinks,
    linkcolor={red!50!black},
    citecolor={blue!50!black},
    urlcolor={blue!80!black}
}
\makeatletter
\renewcommand*\aa@pageof{, page \thepage{} of \pageref*{LastPage}}
\makeatother

\bibpunct{(}{)}{;}{a}{}{,}

\newcommand{\new}[1]{{#1}}
\newcommand{\newbis}[1]{{#1}}
\newcommand{\newthree}[1]{{#1}}

\begin{document}

\title{Multi-epoch scattered-light analysis of HD~135344B: \\ new evidence for a spiral-driving protoplanet}
\titlerunning{Multi-epoch scattered-light analysis of HD~135344B}

\authorrunning{J.~Latour et al.}
\author{J.~Latour\inst{\ref{STARLiege}}\fnmsep\thanks{F.R.S.-FNRS Research Fellow} \and
V.~Christiaens\inst{\ref{KULeuven},\ref{STARLiege}} \and
O.~Absil\inst{\ref{STARLiege}}\fnmsep\thanks{F.R.S.-FNRS Research Director} % continuous feedback
\and
M.~Bonse\inst{\ref{ESOGarching}, \ref{MaxPlanck2}}
\and
R.~Savonet\inst{\ref{STARLiege}}
\and
S.~Juillard\inst{\ref{UArizona}}
\and
I.~Hammond\inst{\ref{MaxPlanck}}
\and
S.~Casassus\inst{\ref{UCH},\ref{DO}}
\and
L.~Cieza\inst{\ref{UDP},\ref{YEMS}}
\and
G.~Cugno\inst{\ref{Zurich}}
\and
C.~Desgrange\inst{\ref{ESOSantiago}}
\and
S.~Lacour\inst{\ref{LIRA}}
\and
D.~Mawet\inst{\ref{CITPasadena},\ref{JPL}}
\and
M.~Montesinos\inst{\ref{UTFChile}}
\and
S.~Perez\inst{\ref{SantiagoChile}, \ref{YEMS}, \ref{CIRAS}}
\and
C.~Pinte\inst{\ref{CNRS}}
\and
M.~Reggiani\inst{\ref{KULeuven}}
\and
T.~Stolker\inst{\ref{Leiden}}
\and
N.~van der Marel\inst{\ref{Leiden}}
\and
A.~Zurlo\inst{\ref{UDP},\ref{YEMS}}}

    \institute{
    STAR Institute, Universit\'e de Li\`ege, All\'ee du Six Ao\^ut 19c, 4000 Li\`ege, Belgium
    \label{STARLiege}
    \and
    Institute of Astronomy, KU Leuven, Celestijnenlaan 200D, Leuven, Belgium
    \label{KULeuven}
    \and
    European Southern Observatory, Karl-Schwarzschild-Straße 2, 85748, Garching bei München, Germany
    \label{ESOGarching}
    \and
    {Max Planck Institute for Intelligent Systems, Max-Planck-Ring 4, D-72076 Tübingen, Germany}
    \label{MaxPlanck2}
    \and
    Steward Observatory, University of Arizona, 933 N. Cherry Avenue, Tucson, AZ 85721, USA
    \label{UArizona}
    \and
    {Max-Planck Institute for Astronomy (MPIA), Königstuhl 17, 69117 Heidelberg, Germany}
    \label{MaxPlanck}
    \and
    Departamento de Astronom\'{\i}a, Universidad de Chile, Casilla 36-D, Santiago, Chile\\
   \email{simon@das.uchile.cl}  \label{UCH}
   %\thanks{Shows the usage of elements in the author field}
   \and
   {Data Observatory Foundation, Eliodoro Y\'a\~{n}ez 2990, Providencia, Santiago, Chile} \label{DO}
   \and
   {Instituto de Estudios Astrofísicos, Facultad de Ingeniería y Ciencias, Universidad Diego Portales, Av. Ejército Libertador 441, Santiago, Chile}
   \label{UDP}
   \and
   {Millennium Nucleus on Young Exoplanets and their Moons (YEMS), Chile.}
   \label{YEMS}
   \and
   {Department of Astrophysics, University of Zurich, Winterthurerstrasse 190, 8057 Zürich, Switzerland}
   \label{Zurich}
   {European Southern Observatory, Alonso de Córdova 3107, Casilla 19, Santiago, Chile}
   \label{ESOSantiago}
   \and
   {LIRA, Observatoire de Paris, Université PSL, Sorbonne Université, Université Paris Cité, CY Cergy Paris Université, CNRS, 92190 Meudon, France}
   \label{LIRA}
   \and{Department of Astronomy, California Institute of Technology, Pasadena, California, USA}
   \label{CITPasadena}
   \and
   {Jet Propulsion Laboratory, California Institute of Technology, Pasadena, California, USA}
   \label{JPL}
   \and
   {Departamento de Física, Universidad Técnica Federico Santa María, Avenida España 1680, Valparaíso, Chile}
   \label{UTFChile}
   \and
   {Departamento de Física, Universidad de Santiago de Chile, Av. V\'ictor Jara 3493, Santiago, Chile} 
   \label{SantiagoChile}
    \and
    {Center for Interdisciplinary Research in Astrophysics Space Exploration (CIRAS), Universidad de Santiago, Chile} 
    \label{CIRAS}
    \and
    {Univ. Grenoble Alpes, CNRS, IPAG, 38000 Grenoble, France}
    \label{CNRS}
    \and
    {Leiden Observatory, Leiden University, P.O. Box 9513, 2300 RA Leiden, The Netherlands}
    \label{Leiden}
    }
    
    \date{Received **, Accepted **}
    
    \abstract
    {The HD 135344B (SAO 206462) disk exhibits strong signposts of planet formation both in scattered light and sub-mm continuum images. ALMA images in the sub-mm revealed a gap-crossing dust filament whose position coincides with a twist detected in the scattered-light spiral structure. Analysis of the spiral dynamics in polarized light also hints at a spiral-driving protoplanet in the sub-mm gap.}
    {We aim to study the overall dynamics of the three spirals in the disk, as well as the motion of the twist over a 10-year baseline, at different IR wavelengths. We also seek to assess the authenticity of a candidate protoplanet recently claimed in the disk.}
    {We use high-fidelity post-processing algorithms such as iterative principal component analysis to minimize the biases induced by angular differential imaging on extended sources and conduct a thorough analysis of archival VLT/NACO, VLT/SPHERE, \new{and} VLT/ERIS datasets in order to obtain the spiral traces and measure their orbital motion in multiple wavelength bands in scattered light. \new{We also reprocess archival JWST/NIRCam datasets with these algorithms.}}
    {We measure an average spiral orbital motion of \newbis{0$\fdg$81 $\pm$ 0$\fdg$05 yr$^{-1}$}, in agreement with the literature value of about 0\fdg85\,yr$^{-1}$ at all wavelengths. With simple modeling of the twist morphology, we confirm that it is indeed co-moving with the spiral in which it is embedded. While the position angle of the twist coincides with the dust filament, it is located at a smaller angular separation from the star, which we attribute to the fact that the spiral trace moves away from the central star with increasing wavelength. We find that a previously claimed protoplanet candidate in the disk can be adequately explained as a post-processing artifact.}
    {\new{Our confirmation that the motion of the scattered light twist is consistent with the orbital velocity of a planet at \newbis{$69\pm 4$} au over a 10-year baseline suggests} that the spirals, the gap and the dust filament in the sub-mm continuum, as well as the twist in scattered light, \new{could} indeed all be attributed to the same hypothetical protoplanet deeply embedded within the spiral. A perplexing trend for a wavelength-dependence of the angular distance of the spiral traces to the central star still remains to be explained.}
   
    \maketitle
    \nolinenumbers

\section{Introduction}
Direct imaging of protoplanetary disks is made possible by state-of-the-art instruments capable of reaching high contrast and high angular resolution on large telescopes. The search for protoplanets embedded within these protoplanetary disks is still in its infancy, as such planets mostly seem to elude detection so far. Only a few of them have been robustly detected in both the IR and H$\alpha$, the well-known PDS~70b and c \citep{Keppler_2018, Haffert_2019}, and WISPIT 2b \citep{close2025, vanCap2025}, although there is no shortage of potential candidates \citep[e.g.,][]{Hammond2023,Currie2025}. One difficulty is that post-processing algorithms tailored to the detection of point-like sources, making use of angular differential imaging \citep[ADI,][]{marois2006}, tend to struggle and bias the final result when applied to extended sources such as protoplanetary disks \citep[e.g.,][]{Milli2012, Juillard2022}. However, recent developments in post-processing algorithms, such as iterative principal component analysis \citep[IPCA, ][]{Stapper2022,Juillard_2024}, enable more unbiased imaging of the birth environments of planets. As a consequence of these recent improvements, revisiting archival datasets can prove valuable to study disk morphology and evolution over time, which is key to apprehending planet formation mechanisms.

HD 135344B (SAO 206462) is a 1.7$^{+0.2}_{-0.1}$ $M_\odot$ \citep{Muller_2011} star located at $135.0\pm0.4$~pc \citep{Gaia2023}. This young star \citep[11.9$^{+3.7}_{-5.8}$ Myr,][]{Garufi2018} is a famous example of a protoplanetary disk exhibiting strong planet formation signposts, such as spiral arms and a central cavity resolved in scattered light \citep{muto2012, garufi2013}, as well as an annular gap in the sub-mm continuum \citep{vdM2016, Cazzoletti2018}. Although alternative explanations exist, a common interpretation for the existence of these structures is the interaction with massive companions \citep{dong2015, price2018}. In the case of HD~135344B, observations specifically point towards the planet-disk interaction scenario. Thanks to measurements of the proper motion of stars from Gaia DR3 \citep{Gaia2023}, recent stellar flybys \citep{Cuello2019} were ruled out as the origin of the spirals \citep{Shuai2022}. While gravitational instability \citep{Lodato_2004} can also be the cause of spirals in protoplanetary disks, the relatively low mass of the HD~135344B disk favors the planet hypothesis \citep{Dong2018}. Yet, the most tantalizing clue for a planet-driven origin resides in the dynamics of the spirals, first explored by \cite{xie2021}, which does not appear consistent with gravitational instability based on spiral rotation rate measurements. Moreover, a subsequent study by \cite{xie2024} finds a $0\fdg85 \pm 0\fdg05$\,yr$^{-1}$ counterclockwise rotational motion. This suggests that the HD~135344B spirals dynamics could be attributed to a driving protoplanet on an orbit that coincides with a gap and a dust filament observed with ALMA in \cite{Casassus_2021}, which they link to a twist in one of the spiral arms in VLT/SPHERE scattered light images, already detected in \cite{Stolker_2016} in polarized light, at the location of the dust filament. This hypothetical spiral-driving companion, if its position coincides with the dust filament found with ALMA, would be embedded within the disk, rendering any direct detection challenging. So far, there has been no direct detection of such a spiral-driving candidate, despite the fact that this effect was measured in multiple protoplanetary disks \citep{Ren_2020, ren2024}.

\begin{table*}[t] 
\begin{center}
\caption{Summary of the scattered-light observations used in this work.} 
\label{tab:datalog}
\begin{tabular}{lcccccccc}
\hline
\hline
Instrument & UT date &Program& Bands & Coronagraph & \makecell[c]{$\epsilon$\\($\arcsec$)} &\makecell[c]{t$_{\rm exp}$\\(s)}& \makecell[c]{$\Delta$PA\\($\degr$)} &Reference star \\
\hline
NACO & 2013-03-23 &090.C-0443(B)& Lp & / &0.83&10676&133& HD 134555 \\ 
 & 2017-05-07 &099.C-0883(A)& Lp &AGPM&1.42&4763&32& / \\ 
 & 2017-05-23 &099.C-0883(B)& Lp &AGPM&1.13&5517&33& / \\ 
 & 2018-06-01 &0101.C-0924(A)& Lp &AGPM&0.40&3705&54& / \\ 
 & 2018-06-02 &0101.C-0924(A)& Lp &AGPM&0.37&3890&128& IRAS 15097-3636 \\ 
 & 2018-06-29 &0101.C-0924(C)& Lp &AGPM&2.48&2593&105& IRAS 15097-3636 \\ 
\hline
SPHERE & 2015-05-14$^{\rm (\newbis{a})}$ &095.C-0298(A)& \makecell[t]{YJH\\K12} & ALC\_YJH\_S &0.50&\makecell[t]{4096 \\4096}&63& HIP 98470 \\
& 2018-03-26 &2100.C-5045(B)& \makecell[t]{YJH\\K12}&/&0.53&\makecell[t]{2688 \\1876}&63& / \\ 
 & 2023-05-11 &109.23HK.001& \makecell[t]{YJH\\K12} &/&0.65&\makecell[t]{1856 \\1232}&12& HD 136533 \\ 
 & 2023-07-23 &109.23HK.001& \makecell[t]{YJH\\K12} &/&0.53&\makecell[t]{1856 \\ 1232}&60& HD 136533 \\ 
 & 2023-07-24&109.23HK.001& \makecell[t]{YJH\\K12} &/&0.63&\makecell[t]{1840 \\1180}&12& HD 136533 \\ 
 & 2024-03-05 &111.24HH.001& \makecell[t]{YJH\\K12} &/&0.35&\makecell[t]{1856 \\1232}&58& HD 136533 \\ 
  & 2024-03-06 &111.24HH.001& \makecell[t]{YJH\\K12} &/&0.58&\makecell[t]{1856 \\1232}&58& HD 136533 \\ 
\hline
\makecell[l]{NIRCam} & 2023-02-16$^{\rm (\newbis{b})}$ &GTO 1179& \makecell{187N+200W \\405N+410M} &/&/&2674&10& / \\
\hline
ERIS & 2024-04-18$^{\rm (\newbis{c})}$ &113.26B2.001& Lp &AGPM &0.52&4963&55& /\\
\hline
\end{tabular}
\end{center}
\addvspace{-0.7em}
{\small%
Notes: The columns provide the instrument used for the observation, the observing night, the program ID, the spectral bands, the coronagraph used, the average seeing, the total integration time (DIT $\times$ N$_{int}$), the amount of field rotation (for total intensity observations), and the name of a reference star if one was used. $^{\rm (\newbis{a})}$ published in \cite{Maire_2017}.  $^{\rm (\newbis{b})}$ published in \cite{cugno2024}. $^{\rm (\newbis{c})}$ published in \cite{maio2025}.
}
\end{table*}

It should be noted that the study of the HD~135344B spiral dynamics is made slightly more complicated by the presence of an unresolved misaligned inner disk \citep{Bohn_2022} that causes shadowing on the outer disk. The shadowed area caused by the inner disk is variable over periods of time as short as a day and can affect the apparent morphology of the outer disk, as observed in \cite{Stolker_2016} and \cite{Stolker_2017}. Additionally, it was shown that both static and moving shadows from a misaligned inner disk can hydrodynamically trigger large-scale spirals \citep{Montesinos2016}, possibly even resembling planetary signatures \citep{Montesinos2018}. However, the spiral pattern orbital period from the shadow-driven scenario is expected to be the same as the inner disk precession period, typically of the order of thousands of years, which the measured velocity of $0\fdg85$\,yr$^{-1}$ in \citet{xie2024} seems to disprove.

A couple of candidate protoplanets around HD~135344B were previously proposed in the literature. Using JWST/NIRCam, \cite{cugno2024} presented a candidate detection outside the spirals, located at a separation of about 300 au, with an estimated mass of 0.8 Jupiter masses. Another candidate at a separation of 28 au, embedded within the protoplanetary disk, was announced in \cite{maio2025} based on VLT/ERIS Lp-band observations of the system.

Here, we aim to investigate the protoplanetary disk of HD 135344B, its morphology and dynamics over time through the use of high-fidelity post-processing algorithms that preserve the morphology of such extended sources, while leaving the study of the central cavity and the search for protoplanets embedded within it for future work (Latour et al. in prep.). Similarly to what \cite{xie2024} did in polarized light, we measure the spiral rotation rate in scattered total intensity light in different wavelength bands, with a longer time baseline, and also analyze the dynamics of the twist. We describe the observations and the data pre-processing in Sect.~\ref{sec:obs}. Then, we present all of our results in Sect.~\ref{sec:results}, and discuss our different interpretations in Sect.~\ref{sec:discussion}.

\section{Observations and image processing} \label{sec:obs}

\subsection{Observations}
For this study, we used archival observations of HD 135344B, captured with VLT/NACO, VLT/SPHERE, JWST/NIRCam, and VLT/ERIS (see Table~\ref{tab:datalog} for all relevant information). Most of these datasets are still unpublished, but some have already been analyzed in the literature. This is the case of the 2015 SPHERE coronagraphic dataset presented in \cite{Maire_2017}, the 2023 JWST non-coronagraphic observation analyzed in \cite{cugno2024}, and, finally, the 2024 ERIS coronagraphic dataset analyzed in \cite{maio2025}. All of the SPHERE observations captured in 2023 and 2024 are non-coronagraphic, non-saturated observations captured in the star-hopping mode \citep{Wahhaj_2021}. This observing mode provides highly correlated reference images of a nearby star that make it possible to use a post-processing algorithm based on Reference Differential Imaging \citep[RDI, ][]{Xie2022}, allowing for a less biased retrieval of the disk morphology. \new{Contrary to all other datasets, the JWST data were not used for spiral motion measurements due to imperfect wavelength match with other observations, which would bias any estimate of spiral motion. Instead, the JWST/NIRCam data are used to both search for accreting embedded protoplanets through the Pa-$\alpha$ (1.87 $\mu$m) and Br-$\alpha$ (4.05 $\mu$m) lines, and illustrate the performances of IPCA in recovering the disk structure at these wavelength. In Fig. \ref{fig:F410M}, we also recover the signal found by \cite{cugno2024} in the F410M filter.}

\subsection{Pre-processing}
The multiple NACO datasets, all of them captured in the Lp band, were calibrated and pre-processed thanks to a pipeline based on PynPoint \citep{Stolker2019}, explained in more detail in a future publication (Bonse et al. in prep). The main steps include classical dark subtraction, flat-fielding, background subtraction, and precise alignment of the science frames.

\begin{figure*}[!t]
    \centering
    \includegraphics[width=\textwidth]{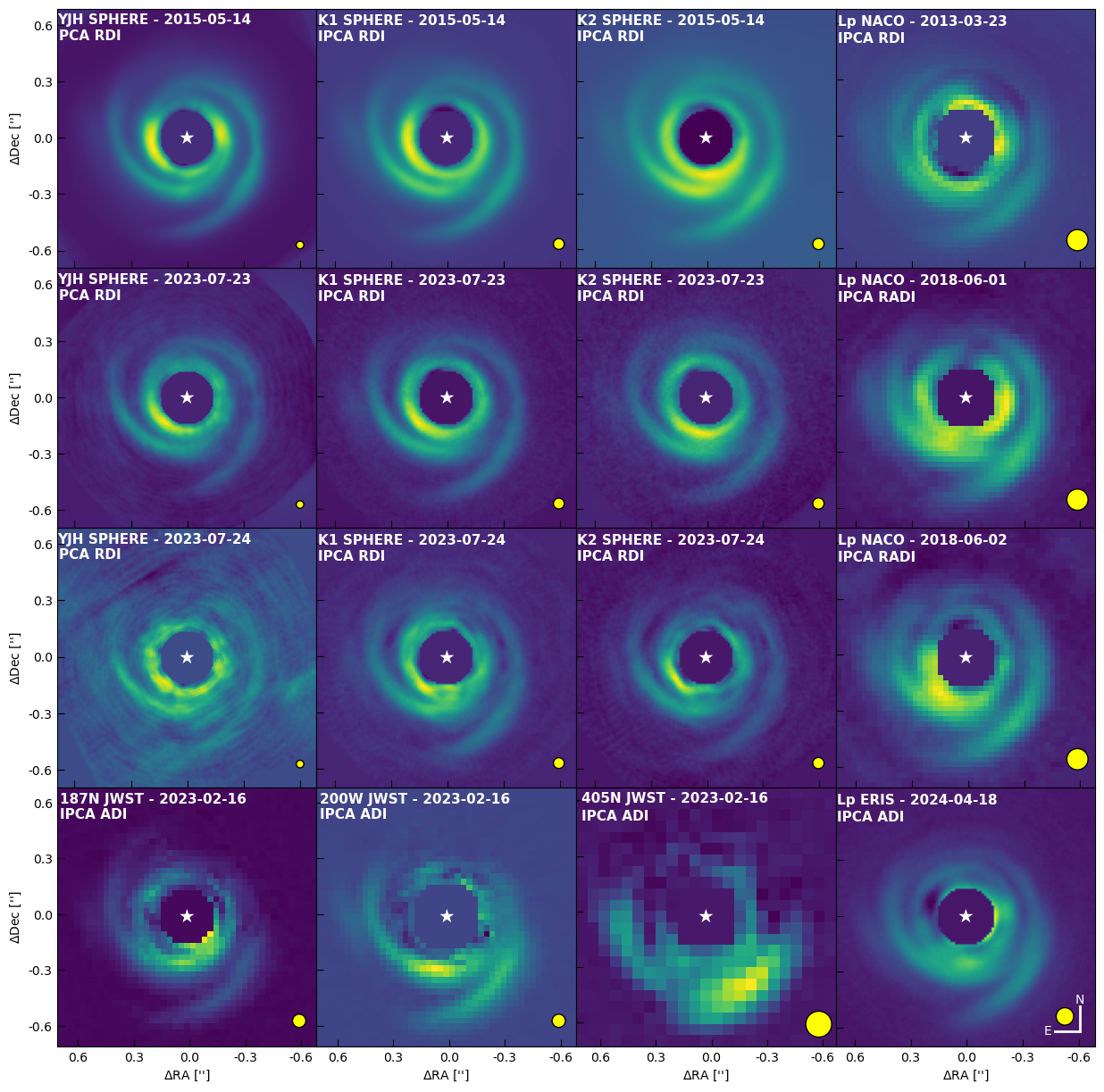}
    \caption{Image gallery of the HD 135344B protoplanetary disk at different wavelengths for multiple datasets. The central star is masked and marked by a white star marker. All of the images were processed using IPCA, except for the images in the YJH bands. North is up, East is left. \new{The yellow circles indicate the spatial resolution of each image.}} 
    \label{fig:DiskGallery}
\end{figure*}

As for the SPHERE observations, the calibration and pre-processing were performed by the SpeCal pipeline \citep{Galicher2018} and the data products retrieved from the SPHERE data center \citep{Delorme2017} for the 2015, 2018 and 2023 datasets, while the vcal-sphere pipeline \citep{christiaens2023b} was used for all the other datasets. Both pipelines allow for the full standard reduction procedure for IFS and IRDIS data, with vcal-sphere further leveraging routines from the Vortex Image Processing package \citep[VIP,][]{Gomez2017, Christiaens2023} for PCA-based sky subtraction, bad pixel correction, FFT-based image alignment, and automatic bad frame removal.

\begin{figure*}[!t]
    \centering
    \includegraphics[width=0.9\textwidth]{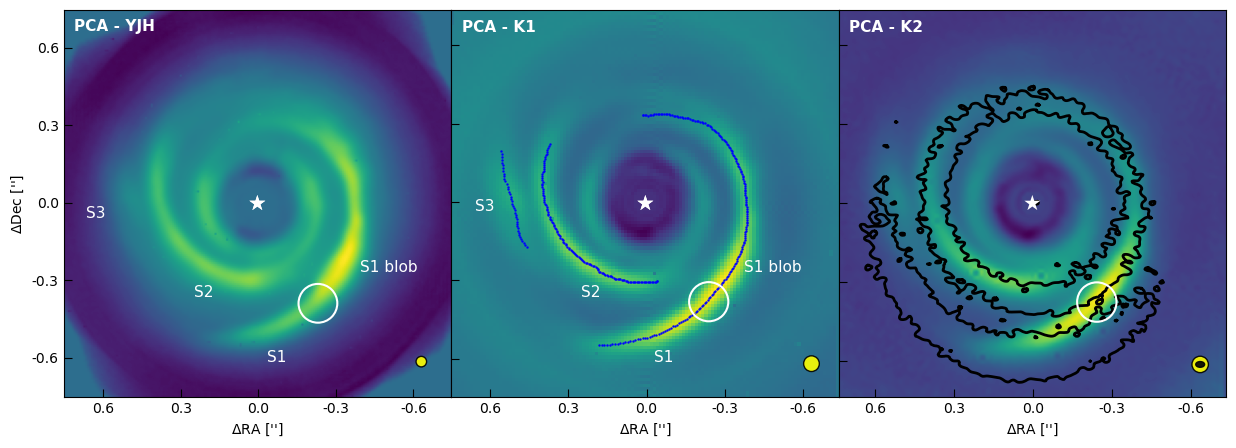}
    \caption{View of the protoplanetary disk of HD 135344B at different wavelengths, post-processing the 2015 coronagraphic SPHERE dataset with PCA-data imputation, which is the dataset that provides the best-quality images of the disk. The different spiral arms are labeled, along with a noticeably brighter region of S1 named the S1 blob. The images were \textit{r}$^{2}$-scaled and deprojected, and the center of the images is marked by the white star. \new{The yellow circles indicate the beam size in the IRDIS data, with the black oval the ALMA beam size.} The white circle marks the position of the twist. The traces extracted from the image in the K1 band are illustrated in blue. Overplotted with the K2 filter are the \new{4-sigma} contours of the ALMA image from \citet{Casassus_2021}, deprojected assuming no flaring since sub-mm continuum observations trace the midplane, and rotated according to the 0.85\degr\,yr$^{-1}$ value found in \citet{xie2024}, highlighting the overlap between the twist and the dust filament.} 
    \label{fig:DiskMorphology}
\end{figure*}

We obtained the calibrated JWST-NIRCam dataset from \cite{cugno2024}. We did not apply any additional pre-processing to this dataset, except for a fine sub-pixel recentering done by maximizing the cross-correlation between the frames on areas of the images that excluded the saturated center of the PSF. 

Finally, the ERIS observation was pre-processed with a custom pipeline that was adapted from a pipeline initially developed for NACO datasets, detailed in \citet{Christiaens2021}, also allowing for dark subtraction, flat-fielding, background subtraction, bad pixel correction, and fine image alignment.

Additionally, we rejected bad frames in every dataset by measuring the correlation of all the frames in the datacube with the median of the cube itself and rejecting a certain percentage of the worst frames. As the PSF was saturated in some datasets, the correlation was calculated on an annulus centered on the PSF core and built in such a way that the saturated core is excluded from the area considered for the calculation. The percentage of rejected frames was determined on a case-by-case basis depending on the overall quality and stability of the dataset, ranging from 0 to a maximum of 5\% of the total images in the datacube.

\subsection{Post-processing}
For all observing nights, the goal was to retrieve the highest-quality image of the disk, i.e., the image that captures the most light from the protoplanetary disk and preserves the best its morphology. To this end, the IPCA \citep{Juillard_2024} algorithm, implemented in VIP is systematically applied to process all of the datasets, as it was shown to preserve protoplanetary disk structures better. \cite{Juillard_2024} compared the use of IPCA with ADI, RDI, or Angular-Reference Differential Imaging (ARDI) and found that the latter generally performs better to retrieve fainter disk signals. However, a fourth strategy, not mentioned in \cite{Juillard_2024}, and named Reference-Angular Differential Imaging (RADI), was sometimes adopted in this paper. It consists in making use of iterative RDI first to retrieve the disk signal as well as possible, and then iterative ADI to better subtract the speckles. This mode is mainly useful for datasets that do not have a lot of field rotation. The best parameters for the IPCA reduction were determined iteratively with a convergence criterion: the flux of the disk retrieved by the algorithm must stabilize and converge after a given number of iterations. For the ERIS dataset, we compare images obtained with PCA and IPCA in order to understand and highlight some artifacts and biases left by algorithms not suited for extended sources. For the 2015 SPHERE observation, PCA-RDI with data imputation \citep{ren_2023} was also used to produce some high-fidelity images of the protoplanetary disk, as it is faster and performs as well as IPCA in this case. Data imputation was not usable for any other dataset, as it requires highly correlated and stable speckles that do not overlap with the disk.

When studying the spiral dynamics, we deprojected the disk assuming a 16\fdg7$\pm$0\fdg6 inclination from face-on \citep{Bohn_2022}. The uncertainty on the inclination was propagated to the deprojected images, but the impact on the spiral traces was found to be negligible. Assumptions on the disk flaring also had to be made for the deprojection. Using polarimetric images, \cite{Avenhaus2018} measured the flaring of multiple disks and found the average value of $\alpha = 1.219 \pm 0.026$ for the flaring index. This range led to a negligible impact on the final deprojected images, even when testing a larger spread of $\alpha$ values than what they suggest. As a consequence, both the uncertainties from the inclination and the disk flaring were neglected. Finally, we also scaled the images with a \textit{r}$^{2}$ map, with \textit{r} the distance to the central star, as the amount of scattered light decreases with the incident light as \textit{r}$^{-2}$. The \textit{r}$^{2}$-scaling, alongside the disk deprojection, is done to avoid geometric biases when tracing the spirals, for which we followed \cite{Ren_2020}.

\section{Results}\label{sec:results}

\subsection{Disk morphology}\label{sec:diskmorphology}
The protoplanetary disk of HD 135344B presents a central cavity depleted of gas and dust, and two large spiral arms. Figure~\ref{fig:DiskGallery} shows a gallery of the most representative post-processed images of HD 135344B in multiple wavelength bands ranging from 1 to 4 microns, using the IPCA algorithm (see Fig.~\ref{fig:FullImageGallery} for the full gallery of all disk images obtained for all of the datasets used in our study). The center of the images is masked to enhance visibility by blocking bright signals and noise in the central cavity, the analysis of which will be presented in Latour et al. (in prep.). Figure~\ref{fig:DiskGallery} suggests that, even when using a pure RDI strategy, it is still better to have more field rotation as illustrated by the lower-quality images for the dataset captured on 2023-07-24, which only presents 12$^\circ$ of rotation. IPCA images of the JWST observations are also presented in Fig.~\ref{fig:DiskGallery} and show a great improvement in the disk recovery over the simple PCA-processed images presented in \cite{cugno2024}. While JWST/NIRCam also observed HD~135344B in the F410M filter, the saturation of the central PSF made it impossible to recover any signal from the protoplanetary disk in that filter; hence why no image is shown using this filter.

All the structures of the disk are better highlighted in $r^{2}$-scaled images and are fully presented and labeled in Fig.~\ref{fig:DiskMorphology}, captured with SPHERE in 2015. The S1 spiral arm takes root North of the star and extends a full 180\degr\ around it out to 0\farcs6. It exhibits a substructure named the S1 blob at a position angle of around 225\degr. At this location, the spiral arm is visibly slightly wider and brighter than its surroundings, which is especially noticeable in the YJH filters. The S2 arm, extending from the South-West to the North-East, appears somewhat smaller and does not have a blob like S1. A much smaller and fainter third spiral arm, referred to as S3, seems to derive from the South-East part of S2, but is not as easily detected at all wavelengths.

\subsection{Spiral orbital motion}\label{sec:spiralmotion}

\begin{figure}[!t]
    \centering
    \includegraphics[width=0.49\textwidth]{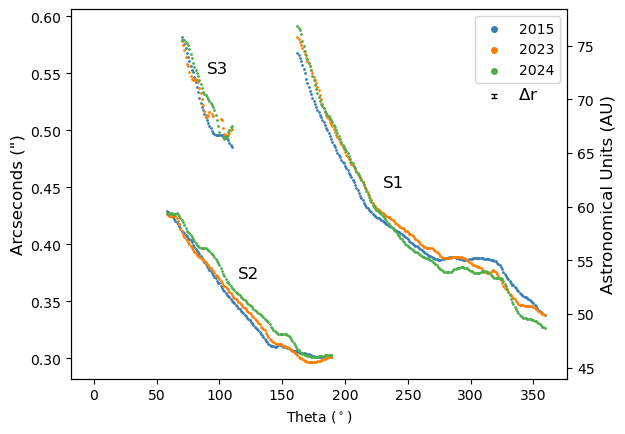}
    \caption{Spiral traces in polar coordinates in the K1 filter in the deprojected images. Theta is the angle measured counterclockwise from North of the central star. \new{In black, labeled $\Delta$r, the typical trace uncertainty, not shown on each data point to improve readability.}}
    \label{fig:SpiralTracesPolar}
\end{figure}

\begin{table*}[t] 
\begin{center}
\caption{Orbital velocity and twist separation measurements.} 
\label{tab:ResultsVelocity}
\begin{tabular}{lccccc}
\hline
\hline
Band & S1 velocity (\degr\ yr$^{-1}$) & S2 velocity (\degr\ yr$^{-1}$) & S3 velocity (\degr\ yr$^{-1}$) & Twist velocity (\degr\ yr$^{-1}$) & Twist separation ($\arcsec$) \\
\hline
YJH & 0.83 $\pm$ 0.14 & 0.69 $\pm$ 0.22 & / & \newbis{1.05 $\pm$ 0.27} & 0.415 $\pm$ \newbis{0.006}\\ 
K1 & 0.80 $\pm$ 0.12 & 0.75 $\pm$ 0.13 & \newbis{0.71 $\pm$ 0.22}& \newbis{0.75 $\pm$ 0.15} & 0.433 $\pm$ \newbis{0.025}\\ 
K2 & 0.94 $\pm$ 0.12 & 0.85 $\pm$ 0.21 & 0.71 $\pm$ 0.19& \newbis{0.70 $\pm$ 0.19} & 0.427 $\pm$ \newbis{0.014}\\
Lp & \newbis{0.83 $\pm$ 0.32} & \newbis{0.80 $\pm$ 0.29} & \newbis{1.18 $\pm$ 0.37}& \newbis{1.18 $\pm$ 0.51} & 0.443 $\pm$ \newbis{0.022} \\ 
\hline
 & 0.86 $\pm$ \newbis{0.07} & 0.77 $\pm$ \newbis{0.09} & \newbis{0.75 $\pm$ 0.14}& \newbis{0.80 $\pm$ 0.11} & 0.419 $\pm$ \newbis{0.005} \\ 
\hline
\end{tabular}
\end{center}
\addvspace{-0.7em}
{\small%
Notes: The columns provide, in order, the orbital velocity measurements for all spiral arms, S1, S2, and S3, as well the twist velocity. The last column is the measured angular separations of the twist. The last line is the weighted average.
}
\end{table*}

We followed the procedure described in \cite{Ren_2020} to study the dynamics of the spirals. Using the deprojected \textit{r}$^2$-scaled images, a Gaussian was fitted to the radial profile of the flux at 1\degr\ intervals in order to obtain the radial maxima for each angle $\theta$. \new{The Gaussian fit was performed on a section of the flux profile centered on the local maximum, of width determined by the distance to the closest local minimum in the flux profile.} The resulting ($r$,$\theta$) pairs for the K1 filter are shown in the middle panel of Fig. \ref{fig:DiskMorphology} and in Fig. \ref{fig:SpiralTracesPolar}. To avoid redundancy, the traces are only shown in the K1 band, but the exact same procedure was repeated for all filters available in our data: YJH, K1, K2, and Lp.

The summary of the results for the spirals velocity is presented in the first three columns of Table~\ref{tab:ResultsVelocity}. Assuming a rigid-body rotation for the spirals, we measured their motion by shifting the traces in polar coordinates relative to one another and finding the angular shift that maximized the cross-correlation between both traces. The angular shift corresponds to the average amount of rotation that each spiral has undergone around the central star in the given time interval between the observations. The \new{$1$-$\sigma$} uncertainties on the measurements were obtained through a classical bootstrapping method with replacement that consisted in creating artificial spiral traces \new{by resampling} the trace data points and obtaining the angular shifts for all these different traces spawned by the original one. The standard deviation of \new{the} values \new{obtained from 1000 bootstrap iterations} was then defined as our uncertainties on our initial measurement. \new{The bootstrapping was done independently on each wavelength band and epoch.} \new{For the K1, K2 and YJH bands, the velocity measurements were always measured relative to the 2015 dataset. This means the angular shift between the 2023 and 2024 datasets for example does not contribute to the final average presented in the table, in order to keep all measurements fully independent of each other. For the Lp band, the measurements were always done relative to the 2013 dataset. The final values for each band, and their associated uncertainties, are obtained through an inverse variance weighted mean. The last line of Tab. \ref{tab:ResultsVelocity} is the weighted average of all bands, with \newbis{the uncertainty of the} the inverse-variance weighted mean.} The velocity of S3 could not be reasonably determined in the YJH band, as this spiral arm \new{has an estimated average SNR below 2 in this band, which is not enough to obtain a proper spiral trace we can trust, in contrast to the K1 and K2 bands where this average SNR is above 5.}

For S1, only the parts of the traces at position angles smaller than 230\degr\ were used for the computation of the angular velocity as this is where the rigid-body assumption seems to hold, as Fig.~\ref{fig:SpiralTracesPolar} shows that the traces intersect with each other at larger position angle values (smaller separations). We conducted a chi-square test on the rigid-body versus Keplerian rotation assumptions for this section of S1 and found reduced chi-square values of 1.96 $\pm$ 0.82 and 4.57 $\pm$ 1.57, respectively. The test thus significantly favors the rigid-body hypothesis. We note that, coincidentally, the twist is included in that section of S1, at the separation of the gap found with ALMA. Additionally, looking closely at Fig.~\ref{fig:DiskMorphology}, we notice a faint prolongation of S2, North of the central star, that seems to extend and merge with S1 in all wavelength bands. The traces we obtain for the position angles between 360\degr\ and 270\degr\ could be influenced by this effect and are therefore not trusted for this exercise.

\new{The uncertainty on the individual trace data points, taking into account the statistical uncertainties of the Gaussian fit, the uncertainty on the inclination of the disk and the flaring index lead to a negligible typical uncertainty of 0.02 pixels \newbis{(0.25 mas in K band)}; hence why no uncertainties are visible in Fig. \ref{fig:SpiralTracesPolar}, as they are too small. The centering uncertainty on our datasets must however also be taken into account, as a small centering offset can lead to systematic displacements of the trace that could be interpreted as motion. The centering procedure for the non-coronagraphic SPHERE datasets was to fit a 2D Gaussian model to the center of the PSF and apply sub-pixel shifts to the frames accordingly. This leads to a typical centering uncertainty of 0.005 pixels \newbis{(0.06 mas in K band)}. Centering the coronagraphic 2015 SPHERE dataset based on speckle cross-correlation leads to a 0.02 pixels uncertainty \newbis{(0.25 mas in K band)}. This uncertainty is trickier to estimate for NACO coronagraphic datasets as the centering of the star behind the coronagraph is not as straightforward. \cite{Godoy2022} cites a ~0.2-pixel uncertainty \newbis{(5.4 mas in Lp band)}. For the ERIS vortex coronagraph, we estimate approximately a 0.1-pixel \newbis{(1.3 mas in Lp band)} pointing uncertainty from \cite{Orban2024}. Assuming a 0.2-pixel systematic shift in the same direction for all the datacube frames, which is the worst case scenario as one could assume those shifts should be roughly equally distributed in all directions, this shift is sufficient to lead to an estimated average measured displacement of the spirals of about 2\degr\ in NACO data. Considering that the angular shifts measured are of the order of 5 to 10\degr\ depending on the datasets, this is significant; however, it remains a high upper bound on the uncertainty due to recentering as the assumption of a systematic shift remaining constant all throughout the observation is unlikely, \newthree{and we assume it is not the case with our observations}. Nevertheless, the Lp results uncertainties should be taken with care. In contrast, with the same assumptions, this average displacement is only of about 0.01\degr\ for SPHERE non-coronagraphic data, and 0.06\degr\ for the sole SPHERE coronagraphic dataset, and 0.3\degr\ for the ERIS data. Additionally, the results in the Lp band also take advantage of two different instruments, NACO and ERIS. Inter-instrument systematics may introduce additional error sources. Yet, using only the NACO datasets, or including ERIS in the measurements, does not change significantly the final values presented, except for the uncertainty that is lowered when adding the ERIS data thanks to the longer time baseline.}

\new{We also estimated the effect of the seeing and, more precisely, the PSF size on the recovered spirals traces, and found it to be negligible. \newbis{Between datasets, the FWHM varies by no more than 5 mas, the most in Lp band, and convolving} our images with PSFs of differing FWHM \newbis{within that range} does not significantly change the recovered traces, \newbis{ with an uncertainty of 0.8 mas on the trace radial location}. The uncertainty on the True North correction must also be taken into account but were deemed negligible except for the NACO instrument. The values are shown in Tab.~\ref{tab:TNCorrections}. Uncertainties on the trace position associated with the post-processing algorithms are difficult to quantify, but our tests suggest that the final images are robust to a wide range of post-processing parameters. The measured dispersion on the trace radial locations is typically inferior to 0.13 pixels \newbis{(1.6 mas in K band)} for different IPCA parameters\newbis{, such as the number of iterations that we varied between 20 and 200, the number of principal components subtracted ranging from 1 to 20, and the reduction strategy changed between ADI, RDI, ARDI or RADI. \newthree{The traces extracted from the two datasets captured one day apart in 2024 differ only by about 0.1 pixels (1.2 mas in K band), which is consistent with the typical uncertainty found for different IPCA parameters.} This stability of the recovered disk with IPCA is in agreement with \cite{Juillard_2024}, who detailed the performance of the IPCA algorithm}. \newthree{Despite its excellent performance for most data sets, there are still some limitations to the use of IPCA, especially for data sets with a low parallactic angle rotation or with only an ADI strategy available, where the disk is not fully recovered (see Fig. \ref{fig:FullImageGallery}). This is the case for the 2017 NACO observations that were not used in our measurements as the disk could not be recovered at all. The 2018 SPHERE dataset, processed with ADI only, is also clearly subject to self-subtraction of the disk; however, the measured dispersion on the trace radial location is the same as for RDI observations, and it is therefore still used in our measurements.} Except for NACO, the internal statistical uncertainties dominate over the uncertainties due to the IPCA reduction \newbis{and other instrumental systematics, as expected from the  instrumental stability of SPHERE over time \citep{Maire2021}.}}

\subsection{Dependence on the wavelength}
\begin{figure}[!t]
    \centering
    \includegraphics[width=0.49\textwidth]{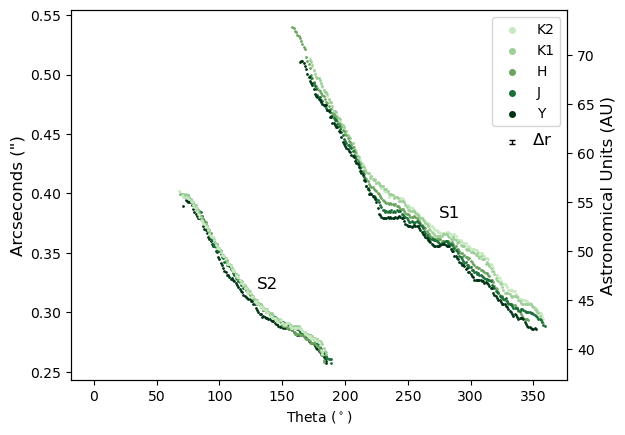}
    \caption{Spirals traces in polar coordinates for the same observation (SPHERE 2015) at different wavelengths. Theta is the angle measured counterclockwise from North of the central star. Contrary to Fig.~\ref{fig:SpiralTracesPolar},  images were not deprojected here. S3 is not included as it was not well detected in all wavelength bands.  \new{In black, labeled $\Delta$r, the typical trace uncertainty, not shown on each data point to improve readability.}}
    \label{fig:SpiralTracesLambda}
\end{figure}

Following a similar procedure, we set out to assess the impact of the wavelength on the trace of the spirals. However, deprojecting the disk requires assumptions for the disk flaring, which is dependent on the wavelength. As such, we opted not to deproject the images for this exercise as any assumption on our part could bias the results. Also, the disk being almost face-on means that any deprojection effect should be minimal and a reasonable trace can be obtained without deprojecting. The results are visible in Fig.~\ref{fig:SpiralTracesLambda}, using the 2015 SPHERE data, which were chosen for illustration as they provide the best quality images of the disk within a great wavelength range. The other datasets show the exact same trends. For both spiral arms, we observe the spirals to be moving away from the central star with increasing wavelength, although this tendency is much stronger for S1 than S2. No proper measurements could be obtained for S3. 

\subsection{S1 twist}

\begin{figure}[!t]
    \centering
    \includegraphics[width=0.49\textwidth]{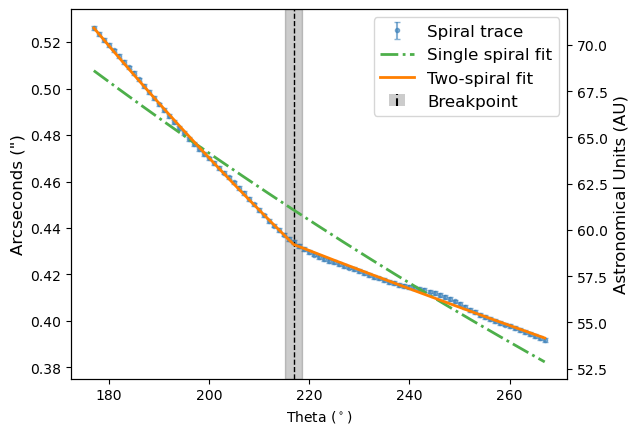}
    \caption{Zoomed in view of the spirals trace for the 2015 SPHERE observation. A single and two-spiral fit are compared. The vertical line shows the breakpoint (the twist) found by the model between the two spirals for the two-spiral fit, along with its uncertainties in the shaded area.}
    \label{fig:TwoSpiralFit2015}
\end{figure}

Here, we investigate the twist that was already observed in scattered light in the structure of S1 \citep{Stolker_2016, Casassus_2021}. In order to do so, we fit a simple logarithmic spiral of equation:
\begin{equation}
r(\theta) = a\exp(b \theta)
    \label{LogSpiral}
\end{equation}
to the trace of the spiral, with \textit{a} and \textit{b} real numbers. Then, we allow for a more complex double-spiral model, in which a breakpoint is built-in. Beyond this breakpoint, the spiral is assumed to follow a logarithmic spiral with parameters that have changed with respect to the first section. This can be summarized as follows:
\begin{equation}
    r(\theta) = 
    \begin{cases}
\text{$a_{1}\exp(b_{1} \theta)$}, & \theta < \theta_{c} \\
\text{$a_{2}\exp(b_{2} \theta)$}, & \theta > \theta_{c}
\end{cases}
\label{DoubleSpiralEquation}
\end{equation}
with $\theta_{c}$ the breakpoint between both spiral models. Continuity was not strictly enforced at $\theta_{c}$. \new{The location of the breakpoint was left entirely as a free parameter in the double-spiral model, bringing the total to five free parameters.} A comparison of the single and the two-spiral fits for the 2015 SPHERE dataset in the K1 filter is shown in Fig.~\ref{fig:TwoSpiralFit2015}. It is clear that the double-spiral model fits the data much better. The model finds the breakpoint at a position angle of $216\fdg58 \pm \newbis{1\fdg65}$. Based on this model, we are then able to track the motion of this structure over time and infer its orbital velocity \new{by independently fitting the two-spiral model to all epochs and wavelengths}. \newbis{Fig. \ref{fig:TwistPosition} shows the position of the twist as a function of time in all wavelength bands, displaying a clear trend for its motion, despite relatively large uncertainties.} The results of this analysis are displayed in the second-to-last column of Table~\ref{tab:ResultsVelocity}. \new{The possibility of a discontinuity was entertained as a way to more closely resemble the model used in \cite{Casassus_2021}. However, enforcing continuity at the breakpoint did not have a great impact on the resulting fit and angular location of the breakpoint. Our observations do not seem to confirm the theoretical framework used in \cite{Casassus_2021} to model the twist in the spiral, as the fit on our data results in a reduced chi square value of 59.9 $\pm$ 6.6 compared to 2.6 $\pm$ 0.2 for our two-spiral model.}

\begin{figure}[!t]
    \centering
    \includegraphics[width=0.47\textwidth]{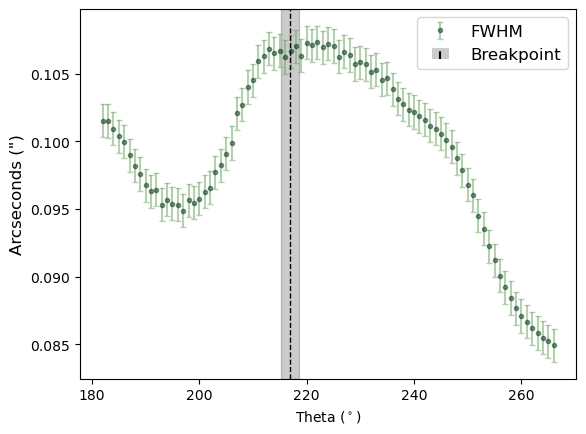}
    \caption{FWHM of S1 in the K1 filter, and measured in the 2015 SPHERE observation. The breakpoint indicating the detected twist is also shown and is located exactly where the spiral arm is wider.}
    \label{fig:FWHM}
\end{figure}

When fitting a Gaussian to the radial profile of a spiral, the centroid of the Gaussian gives us the \new{spiral} trace, but we also retrieved the full-width at half maximum (FWHM), which can then be interpreted as a metric for the width of the spiral. For S1, a clear correlation can be observed between the spiral arm width and the twist, as the latter is always enclosed, or at the boundary, of a wider section of the arm, as illustrated in Fig.~\ref{fig:FWHM}. This wider section of the arm can be recognized as the S1 blob, as named in Fig.~\ref{fig:DiskMorphology} and already identified in \cite{Maire_2017}. This blob also generally looks brighter than its surroundings. \newthree{Additional measurements are included in Fig.~\ref{fig:BlobPosition} and show the S1 blob in multiple epochs and wavelengths. While our data show that the twist position is coincident with the S1 blob, it does not allow us to definitely link the twist to the S1 blob as we are unable to measure reliably the motion of this morphological feature and confirm that it is co-moving with the breakpoint that we measure. A close-up view of the twist with the blob over multiple epochs is also available in Fig. \ref{fig:TwistVisualization}.}

\section{Discussion} \label{sec:discussion}

\subsection{Spiral-driving protoplanet}\label{sec:SpiralDriving}
Multiple elements in our results, along with previous works on this protoplanetary disk, hint at the existence of a spiral-driving protoplanet that remains undetected so far.

First, our orbital velocity measurements presented in Table \ref{tab:ResultsVelocity} are in general agreement with the value of $0\fdg85\pm0\fdg05$\,yr$^{-1}$ found by \cite{xie2024} in polarized light. To find this value, they assumed that the arms are co-rotating, as their measurements seemed in agreement with that hypothesis. \new{Here}, we fitted the orbital velocity of each spiral arm separately, at multiple wavelengths in scattered light. While our larger uncertainties do not allow us to rule out that the spiral arms are indeed co-rotating, there seems to be a trend for S2 to rotate slightly slower. Nevertheless, the orbital velocity we found for S1 is perfectly consistent with the measurement of \cite{xie2024} for all wavelengths tested. If we assume that the spiral arms and the twist \newbis{are} co-rotating, our measurements in Table~\ref{tab:ResultsVelocity} yield an average orbital velocity of \newbis{0$\fdg$81 $\pm$ 0$\fdg$05} yr$^{-1}$, \new{\newbis{with the} inverse-variance weighted mean uncertainty}. This \new{is in agreement with} their prediction for a spiral-driving protoplanet at a radius of about $0\farcs49 \pm 0\farcs02$ (i.e. $66\pm3$\,au), assuming a circular orbit with the same period as the spiral arms. \newbis{Our value results in a semi-major axis estimate of $69\pm4$\,au.}

Additionally, \cite{Casassus_2021} noticed, with ALMA, the presence of a dust filament that bridges an annular gap. They also \new{noted that the position of the twist seen in scattered light in \cite{Stolker_2016}} roughly coincides with this dust filament. This dust filament and twist overlap in terms of orbital distance with the protoplanet suggested by the spiral velocity measurements of \citet{xie2024}. Here, we detect this twist again, at multiple wavelengths, and \new{find an orbital motion that is consistent with it} comoving with the rest of the spiral arm in terms of angular velocity (see Table~\ref{tab:ResultsVelocity}). This seems to indicate that the twist is indeed linked to the same phenomenon that is causing the spiral arms, the annular gap, and the dust filament in the sub-mm continuum all at once; all of which \new{may }point to the existence of a protoplanet, \new{even though other explanations, like multiple perturbers, or more complex spiral dynamics could potentially also explain the same observations, and the positional coincidence of the twist with the dust filament is not definite proof that they are both physically linked}. The twist can also be linked to a blob found in S1 as well, whose position follows the rest of the spiral arm. We also noticed that the twist appears closer to the central star than what is suggested by the spiral-driving orbit found by \cite{xie2024}. Indeed, they cite a separation of $0\farcs49 \pm 0\farcs02$, while we find the twist at about $0\farcs420 \pm 0\farcs011$ (see last column of Table~\ref{tab:ResultsVelocity}). While it may be that the orbit is not circular, we posit that this has more to do with the dependence of the spiral trace on the wavelength of observation, as different wavelengths probe different layers of the protoplanetary disk, as discussed in Sect.~\ref{SpectralSignature}.

We attempted to directly detect a point-like source for this hypothetical protoplanet but were unsuccessful. The SPHERE data from 2015 offer the highest-quality images of the disk but do not present any sign of a point-source in the disk, as already noted by \cite{Maire_2017}. We initially hoped that the JWST-NIRCam observation could shine a new light on this hypothetical protoplanet, but no point source is detected with IPCA, as shown in Fig. \ref{fig:DiskGallery}, despite the filter F187N being centered on Paschen-$\alpha$ and F405N on Brackett-$\alpha$ that could be emitted by accreting giant planets \citep{Aoyama2018}. Nevertheless, it can be noticed that the F405N image presents a brighter region South-East of the central star, roughly at the position of the twist; however, we do not consider this robust enough to claim a detection. Astrometry measurements provide a radial separation of 0$\farcs$45 $\pm$ 0$\farcs$04 and a position angle of 213$\fdg$4 $\pm$ 5$\fdg$7, which is consistent with the twist position. We also obtained a rough estimate of \new{the apparent excess flux on top of the flux from the spiral arm} in the \new{F}405N filter: 0.5$_{-0.3}^{+0.7}$mJy.

\subsection{Spectral signature}\label{SpectralSignature}

Figure \ref{fig:SpiralTracesLambda} shows the dependence of the spiral traces on the wavelength. The traces appear to move away from the central star as the wavelength increases, as the angular separation of the twist in different bands seems to confirm in Table~\ref{tab:ResultsVelocity}. A similar effect was already observed in Table 1 of \cite{Zhong2024} on the RX J1604.3-2130 disk, which is in an almost face-on orientation like HD 135344B. Using 3-D hydrodynamical simulations, \cite{Zhu_2015} studied the structure of spiral shocks induced by young planetary-mass companions and found that the shocks curl inwards towards the central star as they get farther away from the disk midplane. Combining this with the fact that different observing wavelengths probe different depths of the protoplanetary disk, this could explain the trend that emerges from Fig.~\ref{fig:SpiralTracesLambda}. The full analysis of this effect is out of the scope of the present study. Yet, a rudimentary argument can be used to highlight some discrepancies between this model and the data. Longer wavelengths probe deeper layers of the disk, which, based on the 3-D structure found by \cite{Zhu_2015}, is indeed consistent with the traces appearing closer to the star at shorter wavelengths. Nevertheless, a simple geometric argument implies that this effect should be stronger at the near side of the disk rather than at the far side. This is the opposite of what we observe, as the effect is visibly less apparent for S2, located at the near side according to \cite{Stolker_2016}.

\subsection{ERIS protoplanet candidate}\label{sec:candidates}

\begin{figure}[t]
    \centering
    \includegraphics[width=0.49\textwidth]{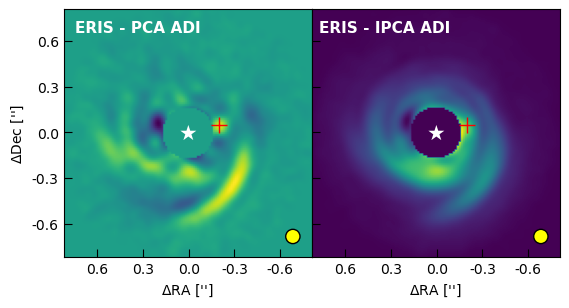}
    \caption{ERIS images processed with PCA (left) and IPCA (right). The center of the images is masked and marked by the white star marker. The candidate found in \cite{maio2025}, detected as they did in the left image with PCA, is marked by the red cross on both images.}
    \label{fig:ERIS-IPCA}
\end{figure}

\cite{maio2025} proposed the detection of a candidate protoplanet located 87\degr\ West of North, at a separation of approximately 0\farcs2, and a contrast of $\sim3 \times 10^{-3}$. We conducted an independent analysis of the same dataset and successfully retrieved the candidate using a PCA-ADI algorithm with a similar number of principal components as they did. However, we also ran the iterative PCA algorithm in an attempt to reduce the geometric biases of ADI on the extended protoplanetary disk, which resulted in the candidate detection blending with the rest of the disk in the final image. The comparison of our PCA and IPCA reductions is shown in Fig.~\ref{fig:ERIS-IPCA}. This highlights how embedded within the disk this candidate would be, and suggests that it may result from the disk being filtered into a point-like source by the ADI algorithm. Additionally, the measured spiral rotation rate suggests that a companion driving the spirals would not be located at that separation.

Besides this re-analysis of the ERIS dataset, we checked all our NACO observations in the Lp band to assess whether this candidate was previously detected, which would reinforce the argument that this is a genuine point-like source. However, the candidate was not robustly detected in any previous datasets. While blobs and overdensities were found in the disk at different positions and epochs, none of these detections were consistent with an object on an Keplerian orbit, rather suggesting a high risk of filtering the spirals and the disk with PCA-ADI. 

\begin{figure}[t]
    \centering
    \includegraphics[width=0.49\textwidth]{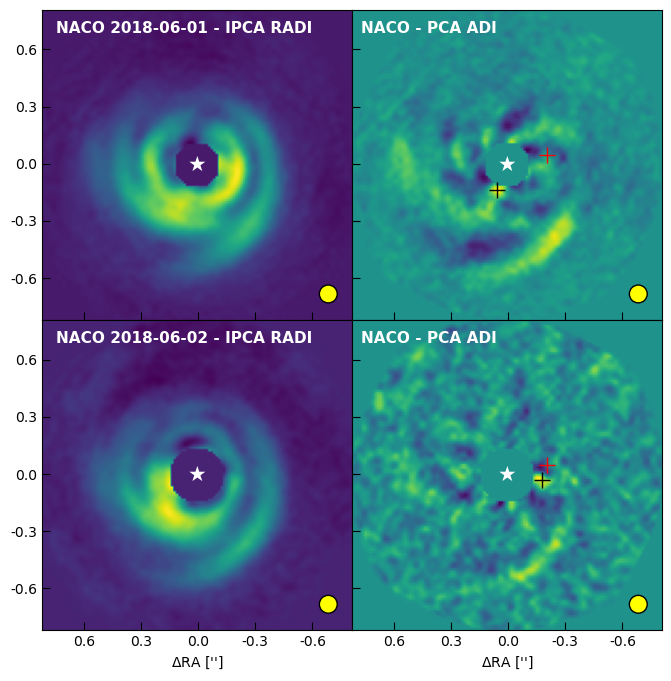}
    \caption{NACO images produced with IPCA (left) from which the disk was extracted to inject in a NACO reference datacube. PCA-processed image of the fake datacube (right). The location of the candidate found by \cite{maio2025} is marked by the red cross, while the black cross marks point-like artifacts generated by the ADI post-processing, which also exhibit what could be interpreted as negative side lobes. \new{The white star marks the center of the images while the yellow circle illustrate the size of the FWHM.}}
    \label{fig:FakeDiskInj}
\end{figure}

In order to test whether the geometric biases of PCA-ADI post-processing can produce point-like sources as observed by \cite{maio2025} for the specific disk morphology of HD 135344B, we ran a test experiment applying PCA-ADI on a simulated datacube containing only a PSF and the protoplanetary disk with no point source within. We started by selecting the highest-quality image of the disk obtained with NACO in the Lp band (epochs 2018-06-01 and 2018-06-02). These disk images were injected into another NACO dataset obtained on a reference PSF, such that we reconstruct an artificial dataset with the same parallactic angle vector as the \citet{maio2025} dataset. The result is displayed in Fig.~\ref{fig:FakeDiskInj} and shows that the geometric biases induced by PCA-ADI can create point-like sources with negative side lobes, similar to the one reported by \citet{maio2025}, which casts additional doubts on this tentative detection.

\section{Conclusions}
We analyzed the HD 135344B protoplanetary disk in scattered light at different near-infrared wavelengths over close to a 10-year baseline. We measured the rotation rate of both spirals and found \newbis{a value of $0\fdg81 \pm 0\fdg05$\,yr$^{-1}$}, consistent with the most recent measurement in \cite{xie2024} of $0\fdg85 \pm 0\fdg05$\,yr$^{-1}$. While our uncertainties do not allow us to reject the hypothesis that the arms are co-rotating, we find a slight trend of S2 moving more slowly than S1.  The hypothesis of a spiral-driving protoplanet, already proposed by \citet{xie2024} \new{is in agreement with} our findings, as we confirmed the presence of a twist in S1 and showed \new{that its motion is consistent with the average motion of the rest of the spiral}, at the orbital velocity expected from \cite{xie2024}. \new{This does not rule out other  hypotheses, such as multiple planetary perturbers and complex disk interactions that may still be able to explain the data.} The position of this twist coincides with the dust filament discussed in \cite{Casassus_2021}. We however do not find any direct hint for a protoplanet at that location in our data sets.

We measured the wavelength dependence of the angular distance of the spiral traces from the central star, noting a perplexing difference in the strength of this effect between both spiral arms, which a pure geometrical argument alone cannot account for.

Lastly, we conducted an independent analysis of the ERIS data presented in \cite{maio2025}. Our analysis suggests that the candidate point-like source presented by \citet{maio2025} is an artifact caused by the use of PCA-ADI processing on the extended protoplanetary disk. We also do not find any sign of this candidate source in any of our other datasets, whether in the Lp band or at other wavelengths, and the measured spiral rotation rate does not indicate the presence of a protoplanet at that separation.

\section*{Data availability}
All disk images presented in Fig. \ref{fig:FullImageGallery} are available at the CDS via anonymous ftp to cdsarc.u-strasbg.fr (130.79.128.5) or via http://cdsweb.u-strasbg.fr/cgi-bin/qcat?J/A+A/.

\begin{acknowledgements}
V.C.\ and O.A.\ thank the Belgian Federal Science Policy Office (BELSPO) for the provision of financial support in the framework of the PRODEX Programme of the European Space Agency (ESA) under contract number 4000142531.
I.H.\ received funding from the European Research Council (ERC) under the European Union’s Horizon 2020 research and innovation programme (PROTOPLANETS, grant agreement No. 101002188).
L.A.C.\ acknowledges support from ANID, FONDECYT Regular grant No. 1241056 and the  Millennium Nucleus on Young Exoplanets and their Moons (YEMS), ANID—Center Code  NCN2024\_001.
Part of this work was carried out within the framework of the NCCR PlanetS supported by the Swiss National Science Foundation under grant 51NF40\_205606.
M.M.\ acknowledges financial support from FONDECYT Regular 1241818.
S.P.\ acknowledges support from FONDECYT grant 1231663 and ANID -- Millennium Science Initiative Program -- Center Code NCN2024\_001.
A.Z.\ acknowledges support from ANID -- Millennium Science Initiative Program -- Center Code NCN2024\_001 and Fondecyt Regular grant number 1250249.

\end{acknowledgements}

\bibliographystyle{aa}
\bibliography{bibliography}

@article{Muller_2011,
   title={HD135344B: a young star has reached its rotational limit},
   volume={530},
   ISSN={1432-0746},
   url={http://dx.doi.org/10.1051/0004-6361/201116732},
   DOI={10.1051/0004-6361/201116732},
   journal={Astronomy \& Astrophysics},
   publisher={EDP Sciences},
   author={Müller, A. and van den Ancker, M. E. and Launhardt, R. and Pott, J. U. and Fedele, D. and Henning, Th.},
   year={2011},
   month=may, pages={A85} }

@ARTICLE{Galicher2018,
       author = {{Galicher}, R. and {Boccaletti}, A. and {Mesa}, D. and {Delorme}, P. and {Gratton}, R. and {Langlois}, M. and {Lagrange}, A.-M. and {Maire}, A.-L. and {Le Coroller}, H. and {Chauvin}, G. and {Biller}, B. and {Cantalloube}, F. and {Janson}, M. and {Lagadec}, E. and {Meunier}, N. and {Vigan}, A. and {Hagelberg}, J. and {Bonnefoy}, M. and {Zurlo}, A. and {Rocha}, S. and {Maurel}, D. and {Jaquet}, M. and {Buey}, T. and {Weber}, L.},
        title = "{Astrometric and photometric accuracies in high contrast imaging: The SPHERE speckle calibration tool (SpeCal)}",
      journal = {\aap},
         year = 2018,
        month = jul,
       volume = {615},
          eid = {A92},
        pages = {A92},
          doi = {10.1051/0004-6361/201832973},
archivePrefix = {arXiv},
       eprint = {1805.04854},
 primaryClass = {astro-ph.IM},
       adsurl = {https://ui.adsabs.harvard.edu/abs/2018A&A...615A..92G}
}

@article{Ren_2020,
   title={Dynamical Evidence of a Spiral Arm–driving Planet in the MWC 758 Protoplanetary Disk},
   volume={898},
   ISSN={2041-8213},
   url={http://dx.doi.org/10.3847/2041-8213/aba43e},
   DOI={10.3847/2041-8213/aba43e},
   number={2},
   journal={The Astrophysical Journal Letters},
   publisher={American Astronomical Society},
   author={Ren , Bin and Dong , Ruobing and van Holstein, Rob G. and Ruffio, Jean-Baptiste and Calvin, Benjamin A. and Girard, Julien H. and Benisty, Myriam and Boccaletti, Anthony and Esposito, Thomas M. and Choquet, Élodie and Mawet, Dimitri and Pueyo, Laurent and Stolker, Tomas and Chiang, Eugene and Boer, Jozua de and Debes, John H. and Garufi, Antonio and Grady, Carol A. and Hines, Dean C. and Maire, Anne-Lise and Ménard, François and Millar-Blanchaer, Maxwell A. and Perrin, Marshall D. and Poteet, Charles A. and Schneider, Glenn},
   year={2020},
   month=jul, pages={L38} }

@article{Juillard_2024,
   title={Combining reference-star and angular differential imaging for high-contrast imaging of extended sources},
   volume={688},
   ISSN={1432-0746},
   url={http://dx.doi.org/10.1051/0004-6361/202449747},
   DOI={10.1051/0004-6361/202449747},
   journal={\aap},
   publisher={EDP Sciences},
   author={Juillard, S. and Christiaens, V. and Absil, O. and Stasevic, S. and Milli, J.},
   year={2024},
   month=aug, pages={A185} }

@article{Keppler_2018,
   title={Discovery of a planetary-mass companion within the gap of the transition disk around PDS 70},
   volume={617},
   ISSN={1432-0746},
   url={http://dx.doi.org/10.1051/0004-6361/201832957},
   DOI={10.1051/0004-6361/201832957},
   journal={\aap},
   publisher={EDP Sciences},
   author={Keppler, M. and Benisty, M. and Müller, A. and Henning, Th. and van Boekel, R. and Cantalloube, F. and Ginski, C. and van Holstein, R. G. and Maire, A.-L. and Pohl, A. and Samland, M. and Avenhaus, H. and Baudino, J.-L. and Boccaletti, A. and de Boer, J. and Bonnefoy, M. and Chauvin, G. and Desidera, S. and Langlois, M. and Lazzoni, C. and Marleau, G.-D. and Mordasini, C. and Pawellek, N. and Stolker, T. and Vigan, A. and Zurlo, A. and Birnstiel, T. and Brandner, W. and Feldt, M. and Flock, M. and Girard, J. and Gratton, R. and Hagelberg, J. and Isella, A. and Janson, M. and Juhasz, A. and Kemmer, J. and Kral, Q. and Lagrange, A.-M. and Launhardt, R. and Matter, A. and Ménard, F. and Milli, J. and Mollière, P. and Olofsson, J. and Pérez, L. and Pinilla, P. and Pinte, C. and Quanz, S. P. and Schmidt, T. and Udry, S. and Wahhaj, Z. and Williams, J. P. and Buenzli, E. and Cudel, M. and Dominik, C. and Galicher, R. and Kasper, M. and Lannier, J. and Mesa, D. and Mouillet, D. and Peretti, S. and Perrot, C. and Salter, G. and Sissa, E. and Wildi, F. and Abe, L. and Antichi, J. and Augereau, J.-C. and Baruffolo, A. and Baudoz, P. and Bazzon, A. and Beuzit, J.-L. and Blanchard, P. and Brems, S. S. and Buey, T. and De Caprio, V. and Carbillet, M. and Carle, M. and Cascone, E. and Cheetham, A. and Claudi, R. and Costille, A. and Delboulbé, A. and Dohlen, K. and Fantinel, D. and Feautrier, P. and Fusco, T. and Giro, E. and Gluck, L. and Gry, C. and Hubin, N. and Hugot, E. and Jaquet, M. and Le Mignant, D. and Llored, M. and Madec, F. and Magnard, Y. and Martinez, P. and Maurel, D. and Meyer, M. and Möller-Nilsson, O. and Moulin, T. and Mugnier, L. and Origné, A. and Pavlov, A. and Perret, D. and Petit, C. and Pragt, J. and Puget, P. and Rabou, P. and Ramos, J. and Rigal, F. and Rochat, S. and Roelfsema, R. and Rousset, G. and Roux, A. and Salasnich, B. and Sauvage, J.-F. and Sevin, A. and Soenke, C. and Stadler, E. and Suarez, M. and Turatto, M. and Weber, L.},
   year={2018},
   month=sep, pages={A44} }

@article{Haffert_2019,
   title={Two accreting protoplanets around the young star PDS 70},
   volume={3},
   ISSN={2397-3366},
   url={http://dx.doi.org/10.1038/s41550-019-0780-5},
   DOI={10.1038/s41550-019-0780-5},
   number={8},
   journal={Nature Astronomy},
   publisher={Springer Science and Business Media LLC},
   author={Haffert, S. Y. and Bohn, A. J. and de Boer, J. and Snellen, I. A. G. and Brinchmann, J. and Girard, J. H. and Keller, C. U. and Bacon, R.},
   year={2019},
   month=jun, pages={749–754} }

@ARTICLE{xie2024,
       author = {{Xie}, Chen and {Xie}, Chengyan and {Ren}, Bin B. and {Benisty}, Myriam and {Ginski}, Christian and {Fang}, Taotao and {Casassus}, Simon and {Bae}, Jaehan and {Facchini}, Stefano and {M{\'e}nard}, Fran{\c{c}}ois and {van Holstein}, Rob G.},
        title = "{Disk Evolution Study Through Imaging of Nearby Young Stars (DESTINYS): Dynamical Evidence of a Spiral-Arm-Driving and Gap-Opening Protoplanet from SAO 206462 Spiral Motion}",
      journal = {Universe},
         year = 2024,
        month = dec,
       volume = {10},
       number = {12},
          eid = {465},
        pages = {465},
          doi = {10.3390/universe10120465},
archivePrefix = {arXiv},
       eprint = {2412.14402},
 primaryClass = {astro-ph.EP},
       adsurl = {https://ui.adsabs.harvard.edu/abs/2024Univ...10..465X}
}

@article{Lodato_2004,
   title={Testing the locality of transport in self-gravitating accretion discs},
   volume={351},
   ISSN={1365-2966},
   url={http://dx.doi.org/10.1111/j.1365-2966.2004.07811.x},
   DOI={10.1111/j.1365-2966.2004.07811.x},
   number={2},
   journal={Monthly Notices of the Royal Astronomical Society},
   publisher={Oxford University Press (OUP)},
   author={Lodato, G. and Rice, W. K. M.},
   year={2004},
   month=jun, pages={630–642} }

@article{Casassus_2021,
   title={A dusty filament and turbulent CO spirals in HD135344B - SAO206462},
   volume={507},
   ISSN={1365-2966},
   url={http://dx.doi.org/10.1093/mnras/stab2359},
   DOI={10.1093/mnras/stab2359},
   number={3},
   journal={Monthly Notices of the Royal Astronomical Society},
   publisher={Oxford University Press (OUP)},
   author={Casassus, Simon and Christiaens, Valentin and Cárcamo, Miguel and Pérez, Sebastián and Weber, Philipp and Ercolano, Barbara and vanderMarel, Nienke and Pinte, Christophe and Dong, Ruobing and Baruteau, Clément and Cieza, Lucas and vanDishoeck, Ewine F and Jordan, Andrés and Price, Daniel J and Absil, Olivier and Arce-Tord, Carla and Faramaz, Virginie and Flores, Christian and Reggiani, Maddalena},
   year={2021},
   month=aug, pages={3789–3809} }

@ARTICLE{cugno2024,
       author = {{Cugno}, Gabriele and {Leisenring}, Jarron and {Wagner}, Kevin R. and {Mullin}, Camryn and {Dong}, Ruobing and {Greene}, Thomas and {Johnstone}, Doug and {Meyer}, Michael R. and {Wolff}, Schuyler G. and {Beichman}, Charles and {Boyer}, Martha and {Horner}, Scott and {Hodapp}, Klaus and {Kelly}, Doug and {McCarthy}, Don and {Roellig}, Thomas and {Rieke}, George and {Rieke}, Marcia and {Stansberry}, John and {Young}, Erick},
        title = "{JWST/NIRCam Imaging of Young Stellar Objects. II. Deep Constraints on Giant Planets and a Planet Candidate Outside of the Spiral Disk Around SAO 206462}",
      journal = {\aj},
         year = 2024,
        month = apr,
       volume = {167},
       number = {4},
          eid = {182},
        pages = {182},
          doi = {10.3847/1538-3881/ad1ffc},
archivePrefix = {arXiv},
       eprint = {2401.02834},
 primaryClass = {astro-ph.EP},
       adsurl = {https://ui.adsabs.harvard.edu/abs/2024AJ....167..182C}
}

@article{Bohn_2022,
   title={Probing inner and outer disk misalignments in transition disks: Constraints from VLTI/GRAVITY and ALMA observations},
   volume={658},
   ISSN={1432-0746},
   url={http://dx.doi.org/10.1051/0004-6361/202142070},
   DOI={10.1051/0004-6361/202142070},
   journal={\aap},
   publisher={EDP Sciences},
   author={Bohn, A. J. and Benisty, M. and Perraut, K. and van der Marel, N. and Wölfer, L. and van Dishoeck, E. F. and Facchini, S. and Manara, C. F. and Teague, R. and Francis, L. and Berger, J.-P. and Garcia-Lopez, R. and Ginski, C. and Henning, T. and Kenworthy, M. and Kraus, S. and Ménard, F. and Mérand, A. and Pérez, L. M.},
   year={2022},
   month=feb, pages={A183} }

@article{Maire_2017,
   title={Testing giant planet formation in the transitional disk of SAO 206462 using deep VLT/SPHERE imaging},
   volume={601},
   ISSN={1432-0746},
   url={http://dx.doi.org/10.1051/0004-6361/201629896},
   DOI={10.1051/0004-6361/201629896},
   journal={\aap},
   publisher={EDP Sciences},
   author={Maire, A.-L. and Stolker, T. and Messina, S. and Müller, A. and Biller, B. A. and Currie, T. and Dominik, C. and Grady, C. A. and Boccaletti, A. and Bonnefoy, M. and Chauvin, G. and Galicher, R. and Millward, M. and Pohl, A. and Brandner, W. and Henning, T. and Lagrange, A.-M. and Langlois, M. and Meyer, M. R. and Quanz, S. P. and Vigan, A. and Zurlo, A. and van Boekel, R. and Buenzli, E. and Buey, T. and Desidera, S. and Feldt, M. and Fusco, T. and Ginski, C. and Giro, E. and Gratton, R. and Hubin, N. and Lannier, J. and Le Mignant, D. and Mesa, D. and Peretti, S. and Perrot, C. and Ramos, J. R. and Salter, G. and Samland, M. and Sissa, E. and Stadler, E. and Thalmann, C. and Udry, S. and Weber, L.},
   year={2017},
   month=may, pages={A134} }

@ARTICLE{maio2025,
       author = {{Maio}, F. and {Fedele}, D. and {Roccatagliata}, V. and {Facchini}, S. and {Lodato}, G. and {Desidera}, S. and {Garufi}, A. and {Mesa}, D. and {Ruzza}, A. and {Toci}, C. and {Testi}, L. and {Zurlo}, A. and {Rosotti}, G.},
        title = "{Unveiling a protoplanet candidate embedded in the HD 135344B disk with VLT/ERIS}",
      journal = {\aap},
         year = 2025,
        month = jul,
       volume = {699},
          eid = {L10},
        pages = {L10},
          doi = {10.1051/0004-6361/202554472},
       adsurl = {https://ui.adsabs.harvard.edu/abs/2025A&A...699L..10M}
}

@article{Wahhaj_2021,
   title={A search for a fifth planet around HR 8799 using the star-hopping RDI technique at VLT/SPHERE},
   volume={648},
   ISSN={1432-0746},
   url={http://dx.doi.org/10.1051/0004-6361/202038794},
   DOI={10.1051/0004-6361/202038794},
   journal={\aap},
   publisher={EDP Sciences},
   author={Wahhaj, Z. and Milli, J. and Romero, C. and Cieza, L. and Zurlo, A. and Vigan, A. and Peña, E. and Valdes, G. and Cantalloube, F. and Girard, J. and Pantoja, B.},
   year={2021},
   month=apr, pages={A26} }

@article{ren_2023,
  title={Karhunen--Lo{\`e}ve data imputation in high-contrast imaging},
  author={Ren, Bin B},
  journal={Astronomy \& Astrophysics},
  volume={679},
  pages={A18},
  year={2023},
  publisher={EDP Sciences}
}

@article{Stolker_2017,
   title={Variable Dynamics in the Inner Disk of HD 135344B Revealed with Multi-epoch Scattered Light Imaging∗},
   volume={849},
   ISSN={1538-4357},
   url={http://dx.doi.org/10.3847/1538-4357/aa886a},
   DOI={10.3847/1538-4357/aa886a},
   number={2},
   journal={The Astrophysical Journal},
   publisher={American Astronomical Society},
   author={Stolker, Tomas and Sitko, Mike and Lazareff, Bernard and Benisty, Myriam and Dominik, Carsten and Waters, Rens and Min, Michiel and Perez, Sebastian and Milli, Julien and Garufi, Antonio and Boer, Jozua de and Ginski, Christian and Kraus, Stefan and Berger, Jean-Philippe and Avenhaus, Henning},
   year={2017},
   month=nov, pages={143} }

@article{Stolker_2016,
   title={Shadows cast on the transition disk of HD 135344B: Multiwavelength VLT/SPHERE polarimetric differential imaging⋆},
   volume={595},
   ISSN={1432-0746},
   url={http://dx.doi.org/10.1051/0004-6361/201528039},
   DOI={10.1051/0004-6361/201528039},
   journal={\aap},
   publisher={EDP Sciences},
   author={Stolker, T. and Dominik, C. and Avenhaus, H. and Min, M. and de Boer, J. and Ginski, C. and Schmid, H. M. and Juhasz, A. and Bazzon, A. and Waters, L. B. F. M. and Garufi, A. and Augereau, J.-C. and Benisty, M. and Boccaletti, A. and Henning, Th. and Langlois, M. and Maire, A.-L. and Ménard, F. and Meyer, M. R. and Pinte, C. and Quanz, S. P. and Thalmann, C. and Beuzit, J.-L. and Carbillet, M. and Costille, A. and Dohlen, K. and Feldt, M. and Gisler, D. and Mouillet, D. and Pavlov, A. and Perret, D. and Petit, C. and Pragt, J. and Rochat, S. and Roelfsema, R. and Salasnich, B. and Soenke, C. and Wildi, F.},
   year={2016},
   month=nov, pages={A113} }

@article{Zhu_2015,
   title={THE STRUCTURE OF SPIRAL SHOCKS EXCITED BY PLANETARY-MASS COMPANIONS},
   volume={813},
   ISSN={1538-4357},
   url={http://dx.doi.org/10.1088/0004-637X/813/2/88},
   DOI={10.1088/0004-637x/813/2/88},
   number={2},
   journal={The Astrophysical Journal},
   publisher={American Astronomical Society},
   author={Zhu, Zhaohuan and Dong, Ruobing and Stone, James M. and Rafikov, Roman R.},
   year={2015},
   month=oct, pages={88} }

@ARTICLE{Gaia2023,
       author = {{Gaia Collaboration} and {Bailer-Jones}, C.~A.~L. and {Teyssier}, D. and {Delchambre}, L. and {Ducourant}, C. and {Garabato}, D. and {Hatzidimitriou}, D. and {Klioner}, S.~A. and {Rimoldini}, L. and {Bellas-Velidis}, I. and {Carballo}, R. and {Carnerero}, M.~I. and {Diener}, C. and {Fouesneau}, M. and {Galluccio}, L. and {Gavras}, P. and {Krone-Martins}, A. and {Raiteri}, C.~M. and {Teixeira}, R. and {Brown}, A.~G.~A. and {Vallenari}, A. and {Prusti}, T. and {de Bruijne}, J.~H.~J. and {Arenou}, F. and {Babusiaux}, C. and {Biermann}, M. and {Creevey}, O.~L. and {Evans}, D.~W. and {Eyer}, L. and {Guerra}, R. and {Hutton}, A. and {Jordi}, C. and {Lammers}, U.~L. and {Lindegren}, L. and {Luri}, X. and {Mignard}, F. and {Panem}, C. and {Pourbaix}, D. and {Randich}, S. and {Sartoretti}, P. and {Soubiran}, C. and {Tanga}, P. and {Walton}, N.~A. and {Bastian}, U. and {Drimmel}, R. and {Jansen}, F. and {Katz}, D. and {Lattanzi}, M.~G. and {van Leeuwen}, F. and {Bakker}, J. and {Cacciari}, C. and {Casta{\~n}eda}, J. and {De Angeli}, F. and {Fabricius}, C. and {Fr{\'e}mat}, Y. and {Guerrier}, A. and {Heiter}, U. and {Masana}, E. and {Messineo}, R. and {Mowlavi}, N. and {Nicolas}, C. and {Nienartowicz}, K. and {Pailler}, F. and {Panuzzo}, P. and {Riclet}, F. and {Roux}, W. and {Seabroke}, G.~M. and {Sordo}, R. and {Th{\'e}venin}, F. and {Gracia-Abril}, G. and {Portell}, J. and {Altmann}, M. and {Andrae}, R. and {Audard}, M. and {Benson}, K. and {Berthier}, J. and {Blomme}, R. and {Burgess}, P.~W. and {Busonero}, D. and {Busso}, G. and {C{\'a}novas}, H. and {Carry}, B. and {Cellino}, A. and {Cheek}, N. and {Clementini}, G. and {Damerdji}, Y. and {Davidson}, M. and {de Teodoro}, P. and {Nu{\~n}ez Campos}, M. and {Dell'Oro}, A. and {Esquej}, P. and {Fern{\'a}ndez-Hern{\'a}ndez}, J. and {Fraile}, E. and {Garc{\'\i}a-Lario}, P. and {Gosset}, E. and {Haigron}, R. and {Halbwachs}, J. -L. and {Hambly}, N.~C. and {Harrison}, D.~L. and {Hern{\'a}ndez}, J. and {Hestroffer}, D. and {Hodgkin}, S.~T. and {Holl}, B. and {Jan{\ss}en}, K. and {Jevardat de Fombelle}, G. and {Jordan}, S. and {Lanzafame}, A.~C. and {L{\"o}ffler}, W. and {Marchal}, O. and {Marrese}, P.~M. and {Moitinho}, A. and {Muinonen}, K. and {Osborne}, P. and {Pancino}, E. and {Pauwels}, T. and {Recio-Blanco}, A. and {Reyl{\'e}}, C. and {Riello}, M. and {Roegiers}, T. and {Rybizki}, J. and {Sarro}, L.~M. and {Siopis}, C. and {Smith}, M. and {Sozzetti}, A. and {Utrilla}, E. and {van Leeuwen}, M. and {Abbas}, U. and {{\'A}brah{\'a}m}, P. and {Abreu Aramburu}, A. and {Aerts}, C. and {Aguado}, J.~J. and {Ajaj}, M. and {Aldea-Montero}, F. and {Altavilla}, G. and {{\'A}lvarez}, M.~A. and {Alves}, J. and {Anderson}, R.~I. and {Anglada Varela}, E. and {Antoja}, T. and {Baines}, D. and {Baker}, S.~G. and {Balaguer-N{\'u}{\~n}ez}, L. and {Balbinot}, E. and {Balog}, Z. and {Barache}, C. and {Barbato}, D. and {Barros}, M. and {Barstow}, M.~A. and {Bartolom{\'e}}, S. and {Bassilana}, J. -L. and {Bauchet}, N. and {Becciani}, U. and {Bellazzini}, M. and {Berihuete}, A. and {Bernet}, M. and {Bertone}, S. and {Bianchi}, L. and {Binnenfeld}, A. and {Blanco-Cuaresma}, S. and {Boch}, T. and {Bombrun}, A. and {Bossini}, D. and {Bouquillon}, S. and {Bragaglia}, A. and {Bramante}, L. and {Breedt}, E. and {Bressan}, A. and {Brouillet}, N. and {Brugaletta}, E. and {Bucciarelli}, B. and {Burlacu}, A. and {Butkevich}, A.~G. and {Buzzi}, R. and {Caffau}, E. and {Cancelliere}, R. and {Cantat-Gaudin}, T. and {Carlucci}, T. and {Carrasco}, J.~M. and {Casamiquela}, L. and {Castellani}, M. and {Castro-Ginard}, A. and {Chaoul}, L. and {Charlot}, P. and {Chemin}, L. and {Chiaramida}, V. and {Chiavassa}, A. and {Chornay}, N. and {Comoretto}, G. and {Contursi}, G. and {Cooper}, W.~J. and {Cornez}, T. and {Cowell}, S. and {Crifo}, F. and {Cropper}, M. and {Crosta}, M. and {Crowley}, C. and {Dafonte}, C. and {Dapergolas}, A. and {David}, P. and {de Laverny}, P.},
        title = "{Gaia Data Release 3. The extragalactic content}",
      journal = {\aap},
         year = 2023,
        month = jun,
       volume = {674},
          eid = {A41},
        pages = {A41},
          doi = {10.1051/0004-6361/202243232},
archivePrefix = {arXiv},
       eprint = {2206.05681},
 primaryClass = {astro-ph.GA},
       adsurl = {https://ui.adsabs.harvard.edu/abs/2023A&A...674A..41G}
}

@ARTICLE{Cuello2019,
       author = {{Cuello}, Nicol{\'a}s and {Dipierro}, Giovanni and {Mentiplay}, Daniel and {Price}, Daniel J. and {Pinte}, Christophe and {Cuadra}, Jorge and {Laibe}, Guillaume and {M{\'e}nard}, Fran{\c{c}}ois and {Poblete}, Pedro P. and {Montesinos}, Mat{\'\i}as},
        title = "{Flybys in protoplanetary discs: I. Gas and dust dynamics}",
      journal = {\mnras},
         year = 2019,
        month = mar,
       volume = {483},
       number = {3},
        pages = {4114-4139},
          doi = {10.1093/mnras/sty3325},
archivePrefix = {arXiv},
       eprint = {1812.00961},
 primaryClass = {astro-ph.EP},
       adsurl = {https://ui.adsabs.harvard.edu/abs/2019MNRAS.483.4114C}
}

@ARTICLE{muto2012,
       author = {{Muto}, T. and {Grady}, C.~A. and {Hashimoto}, J. and {Fukagawa}, M. and {Hornbeck}, J.~B. and {Sitko}, M. and {Russell}, R. and {Werren}, C. and {Cur{\'e}}, M. and {Currie}, T. and {Ohashi}, N. and {Okamoto}, Y. and {Momose}, M. and {Honda}, M. and {Inutsuka}, S. and {Takeuchi}, T. and {Dong}, R. and {Abe}, L. and {Brandner}, W. and {Brandt}, T. and {Carson}, J. and {Egner}, S. and {Feldt}, M. and {Fukue}, T. and {Goto}, M. and {Guyon}, O. and {Hayano}, Y. and {Hayashi}, M. and {Hayashi}, S. and {Henning}, T. and {Hodapp}, K.~W. and {Ishii}, M. and {Iye}, M. and {Janson}, M. and {Kandori}, R. and {Knapp}, G.~R. and {Kudo}, T. and {Kusakabe}, N. and {Kuzuhara}, M. and {Matsuo}, T. and {Mayama}, S. and {McElwain}, M.~W. and {Miyama}, S. and {Morino}, J. -I. and {Moro-Martin}, A. and {Nishimura}, T. and {Pyo}, T. -S. and {Serabyn}, E. and {Suto}, H. and {Suzuki}, R. and {Takami}, M. and {Takato}, N. and {Terada}, H. and {Thalmann}, C. and {Tomono}, D. and {Turner}, E.~L. and {Watanabe}, M. and {Wisniewski}, J.~P. and {Yamada}, T. and {Takami}, H. and {Usuda}, T. and {Tamura}, M.},
        title = "{Discovery of Small-scale Spiral Structures in the Disk of SAO 206462 (HD 135344B): Implications for the Physical State of the Disk from Spiral Density Wave Theory}",
      journal = {\apjl},
         year = 2012,
        month = apr,
       volume = {748},
       number = {2},
          eid = {L22},
        pages = {L22},
          doi = {10.1088/2041-8205/748/2/L22},
archivePrefix = {arXiv},
       eprint = {1202.6139},
 primaryClass = {astro-ph.EP},
       adsurl = {https://ui.adsabs.harvard.edu/abs/2012ApJ...748L..22M}
}

@ARTICLE{garufi2013,
       author = {{Garufi}, A. and {Quanz}, S.~P. and {Avenhaus}, H. and {Buenzli}, E. and {Dominik}, C. and {Meru}, F. and {Meyer}, M.~R. and {Pinilla}, P. and {Schmid}, H.~M. and {Wolf}, S.},
        title = "{Small vs. large dust grains in transitional disks: do different cavity sizes indicate a planet?. SAO 206462 (HD 135344B) in polarized light with VLT/NACO}",
      journal = {\aap},
         year = 2013,
        month = dec,
       volume = {560},
          eid = {A105},
        pages = {A105},
          doi = {10.1051/0004-6361/201322429},
archivePrefix = {arXiv},
       eprint = {1311.4195},
 primaryClass = {astro-ph.EP},
       adsurl = {https://ui.adsabs.harvard.edu/abs/2013A&A...560A.105G}
}

@ARTICLE{marois2006,
       author = {{Marois}, Christian and {Lafreni{\`e}re}, David and {Doyon}, Ren{\'e} and {Macintosh}, Bruce and {Nadeau}, Daniel},
        title = "{Angular Differential Imaging: A Powerful High-Contrast Imaging Technique}",
      journal = {\apj},
         year = 2006,
        month = apr,
       volume = {641},
       number = {1},
        pages = {556-564},
          doi = {10.1086/500401},
archivePrefix = {arXiv},
       eprint = {astro-ph/0512335},
 primaryClass = {astro-ph},
       adsurl = {https://ui.adsabs.harvard.edu/abs/2006ApJ...641..556M}
}

@ARTICLE{xie2021,
       author = {{Xie}, Chengyan and {Ren}, Bin and {Dong}, Ruobing and {Pueyo}, Laurent and {Ruffio}, Jean-Baptiste and {Fang}, Taotao and {Mawet}, Dimitri and {Stolker}, Tomas},
        title = "{Spiral Arm Pattern Motion in the SAO 206462 Protoplanetary Disk}",
      journal = {\apjl},
         year = 2021,
        month = jan,
       volume = {906},
       number = {2},
          eid = {L9},
        pages = {L9},
          doi = {10.3847/2041-8213/abd241},
archivePrefix = {arXiv},
       eprint = {2012.05242},
 primaryClass = {astro-ph.EP},
       adsurl = {https://ui.adsabs.harvard.edu/abs/2021ApJ...906L...9X}
}

@ARTICLE{ren2024,
       author = {{Ren}, Bin B. and {Xie}, Chen and {Benisty}, Myriam and {Dong}, Ruobing and {Bae}, Jaehan and {Stolker}, Tomas and {van Holstein}, Rob G. and {Debes}, John H. and {Garufi}, Antonio and {Ginski}, Christian and {Kraus}, Stefan},
        title = "{A companion in V1247 Ori supported by motion in the pattern of the spiral arm}",
      journal = {\aap},
         year = 2024,
        month = jan,
       volume = {681},
          eid = {L2},
        pages = {L2},
          doi = {10.1051/0004-6361/202348114},
archivePrefix = {arXiv},
       eprint = {2310.15430},
 primaryClass = {astro-ph.EP},
       adsurl = {https://ui.adsabs.harvard.edu/abs/2024A&A...681L...2R}
}

@ARTICLE{Gomez2017,
       author = {{Gomez Gonzalez}, Carlos Alberto and {Wertz}, Olivier and {Absil}, Olivier and {Christiaens}, Valentin and {Defr{\`e}re}, Denis and {Mawet}, Dimitri and {Milli}, Julien and {Absil}, Pierre-Antoine and {Van Droogenbroeck}, Marc and {Cantalloube}, Faustine and {Hinz}, Philip M. and {Skemer}, Andrew J. and {Karlsson}, Mikael and {Surdej}, Jean},
        title = "{VIP: Vortex Image Processing Package for High-contrast Direct Imaging}",
      journal = {\aj},
         year = 2017,
        month = jul,
       volume = {154},
       number = {1},
          eid = {7},
        pages = {7},
          doi = {10.3847/1538-3881/aa73d7},
archivePrefix = {arXiv},
       eprint = {1705.06184},
 primaryClass = {astro-ph.IM},
       adsurl = {https://ui.adsabs.harvard.edu/abs/2017AJ....154....7G}
}

@ARTICLE{Christiaens2023,
       author = {{Christiaens}, Valentin and {Gonzalez}, Carlos and {Farkas}, Ralf and {Dahlqvist}, Carl-Henrik and {Nasedkin}, Evert and {Milli}, Julien and {Absil}, Olivier and {Ngo}, Henry and {Cantero}, Carles and {Rainot}, Alan and {Hammond}, Iain and {Bonse}, Markus and {Cantalloube}, Faustine and {Vigan}, Arthur and {Kompella}, Vijay and {Hancock}, Paul},
        title = "{VIP: A Python package for high-contrast imaging}",
      journal = {The Journal of Open Source Software},
         year = 2023,
        month = jan,
       volume = {8},
       number = {81},
          eid = {4774},
        pages = {4774},
          doi = {10.21105/joss.04774},
       adsurl = {https://ui.adsabs.harvard.edu/abs/2023JOSS....8.4774C}
}

@ARTICLE{close2025,
       author = {{Close}, Laird M. and {van Capelleveen}, Richelle F. and {Weible}, Gabriel and {Wagner}, Kevin and {Haffert}, Sebastiaan Y. and {Males}, Jared R. and {Ilyin}, Ilya and {Kenworthy}, Matthew A. and {Li}, Jialin and {Long}, Joseph D. and {Ertel}, Steve and {Ginski}, Christian and {Weinberger}, Alycia J. and {Follette}, Kate and {Liberman}, Joshua and {Twitchell}, Katie and {Johnson}, Parker and {Kueny}, Jay and {Apai}, Daniel and {Doyon}, Rene and {Foster}, Warren and {Gasho}, Victor and {Van Gorkom}, Kyle and {Guyon}, Olivier and {Kautz}, Maggie Y. and {McLeod}, Avalon and {McEwen}, Eden and {Pearce}, Logan and {Schatz}, Lauren and {Hedglen}, Alexander D. and {Wu}, Ya-Lin and {Isbell}, Jacob and {Power}, Jenny and {Carlson}, Jared and {Close}, Emmeline and {Tonucci}, Elena and {Mars}, Matthijs},
        title = "{Wide Separation Planets in Time (WISPIT): Discovery of a Gap H{\ensuremath{\alpha}} Protoplanet WISPIT 2b with MagAO-X}",
      journal = {\apjl},
         year = 2025,
        month = sep,
       volume = {990},
       number = {1},
          eid = {L9},
        pages = {L9},
          doi = {10.3847/2041-8213/adf7a5},
archivePrefix = {arXiv},
       eprint = {2508.19046},
 primaryClass = {astro-ph.EP},
       adsurl = {https://ui.adsabs.harvard.edu/abs/2025ApJ...990L...9C}
}

@ARTICLE{Hammond2023,
       author = {{Hammond}, Iain and {Christiaens}, Valentin and {Price}, Daniel J. and {Toci}, Claudia and {Pinte}, Christophe and {Juillard}, Sandrine and {Garg}, Himanshi},
        title = "{Confirmation and Keplerian motion of the gap-carving protoplanet HD 169142 b}",
      journal = {\mnras},
         year = 2023,
        month = jun,
       volume = {522},
       number = {1},
        pages = {L51-L55},
          doi = {10.1093/mnrasl/slad027},
archivePrefix = {arXiv},
       eprint = {2302.11302},
 primaryClass = {astro-ph.EP},
       adsurl = {https://ui.adsabs.harvard.edu/abs/2023MNRAS.522L..51H}
}

@ARTICLE{Currie2025,
       author = {{Currie}, Thayne and {Hashimoto}, Jun and {Aoyama}, Yuhiko and {Dong}, Ruobing and {Fukagawa}, Misato and {Muto}, Takayuki and {Dykes}, Erica and {El Morsy}, Mona and {Tamura}, Motohide},
        title = "{VLT/MUSE Detection of the AB Aurigae b Protoplanet with H$_{{\ensuremath{\alpha}}}$ Spectroscopy}",
      journal = {\apjl},
         year = 2025,
        month = sep,
       volume = {990},
       number = {2},
          eid = {L42},
        pages = {L42},
          doi = {10.3847/2041-8213/adf7a0},
archivePrefix = {arXiv},
       eprint = {2508.18351},
 primaryClass = {astro-ph.EP},
       adsurl = {https://ui.adsabs.harvard.edu/abs/2025ApJ...990L..42C}
}

@ARTICLE{Avenhaus2018,
       author = {{Avenhaus}, Henning and {Quanz}, Sascha P. and {Garufi}, Antonio and {Perez}, Sebastian and {Casassus}, Simon and {Pinte}, Christophe and {Bertrang}, Gesa H. -M. and {Caceres}, Claudio and {Benisty}, Myriam and {Dominik}, Carsten},
        title = "{Disks around T Tauri Stars with SPHERE (DARTTS-S). I. SPHERE/IRDIS Polarimetric Imaging of Eight Prominent T Tauri Disks}",
      journal = {\apj},
         year = 2018,
        month = aug,
       volume = {863},
       number = {1},
          eid = {44},
        pages = {44},
          doi = {10.3847/1538-4357/aab846},
archivePrefix = {arXiv},
       eprint = {1803.10882},
 primaryClass = {astro-ph.SR},
       adsurl = {https://ui.adsabs.harvard.edu/abs/2018ApJ...863...44A}
}

@ARTICLE{Zhong2024,
       author = {{Zhong}, Huisheng and {Ren}, Bin B. and {Ma}, Bo and {Xie}, Chen and {Ma}, Jie and {Wallack}, Nicole L. and {Mawet}, Dimitri and {Ruane}, Garreth},
        title = "{Multiband reflectance and shadowing of the protoplanetary disk RX J1604.3-2130 in scattered light}",
      journal = {\aap},
         year = 2024,
        month = apr,
       volume = {684},
          eid = {A168},
        pages = {A168},
          doi = {10.1051/0004-6361/202348874},
archivePrefix = {arXiv},
       eprint = {2402.16698},
 primaryClass = {astro-ph.EP},
       adsurl = {https://ui.adsabs.harvard.edu/abs/2024A&A...684A.168Z}
}

@ARTICLE{Shuai2022,
       author = {{Shuai}, Linling and {Ren}, Bin B. and {Dong}, Ruobing and {Zhou}, Xingyu and {Pueyo}, Laurent and {De Rosa}, Robert J. and {Fang}, Taotao and {Mawet}, Dimitri},
        title = "{Stellar Flyby Analysis for Spiral Arm Hosts with Gaia DR3}",
      journal = {\apjs},
         year = 2022,
        month = dec,
       volume = {263},
       number = {2},
          eid = {31},
        pages = {31},
          doi = {10.3847/1538-4365/ac98fd},
archivePrefix = {arXiv},
       eprint = {2210.03725},
 primaryClass = {astro-ph.EP},
       adsurl = {https://ui.adsabs.harvard.edu/abs/2022ApJS..263...31S}
}

@ARTICLE{Milli2012,
       author = {{Milli}, J. and {Mouillet}, D. and {Lagrange}, A. -M. and {Boccaletti}, A. and {Mawet}, D. and {Chauvin}, G. and {Bonnefoy}, M.},
        title = "{Impact of angular differential imaging on circumstellar disk images}",
      journal = {\aap},
         year = 2012,
        month = sep,
       volume = {545},
          eid = {A111},
        pages = {A111},
          doi = {10.1051/0004-6361/201219687},
archivePrefix = {arXiv},
       eprint = {1207.5909},
 primaryClass = {astro-ph.EP},
       adsurl = {https://ui.adsabs.harvard.edu/abs/2012A&A...545A.111M}
}

@ARTICLE{Juillard2022,
       author = {{Juillard}, S. and {Christiaens}, V. and {Absil}, O.},
        title = "{Analysis of the arm-like structure in the outer disk of PDS 70. Spiral density wave or vortex?}",
      journal = {\aap},
         year = 2022,
        month = dec,
       volume = {668},
          eid = {A125},
        pages = {A125},
          doi = {10.1051/0004-6361/202244402},
archivePrefix = {arXiv},
       eprint = {2211.03361},
 primaryClass = {astro-ph.EP},
       adsurl = {https://ui.adsabs.harvard.edu/abs/2022A&A...668A.125J}
}

@ARTICLE{dong2015,
       author = {{Dong}, Ruobing and {Zhu}, Zhaohuan and {Rafikov}, Roman R. and {Stone}, James M.},
        title = "{Observational Signatures of Planets in Protoplanetary Disks: Spiral Arms Observed in Scattered Light Imaging Can be Induced by Planets}",
      journal = {\apjl},
         year = 2015,
        month = aug,
       volume = {809},
       number = {1},
          eid = {L5},
        pages = {L5},
          doi = {10.1088/2041-8205/809/1/L5},
archivePrefix = {arXiv},
       eprint = {1507.03596},
 primaryClass = {astro-ph.EP},
       adsurl = {https://ui.adsabs.harvard.edu/abs/2015ApJ...809L...5D}
}

@ARTICLE{price2018,
       author = {{Price}, Daniel J. and {Cuello}, Nicol{\'a}s and {Pinte}, Christophe and {Mentiplay}, Daniel and {Casassus}, Simon and {Christiaens}, Valentin and {Kennedy}, Grant M. and {Cuadra}, Jorge and {Sebastian Perez}, M. and {Marino}, Sebastian and {Armitage}, Philip J. and {Zurlo}, Alice and {Juhasz}, Attila and {Ragusa}, Enrico and {Laibe}, Guillaume and {Lodato}, Giuseppe},
        title = "{Circumbinary, not transitional: on the spiral arms, cavity, shadows, fast radial flows, streamers, and horseshoe in the HD 142527 disc}",
      journal = {\mnras},
         year = 2018,
        month = jun,
       volume = {477},
       number = {1},
        pages = {1270-1284},
          doi = {10.1093/mnras/sty647},
archivePrefix = {arXiv},
       eprint = {1803.02484},
 primaryClass = {astro-ph.SR},
       adsurl = {https://ui.adsabs.harvard.edu/abs/2018MNRAS.477.1270P}
}

@ARTICLE{Stolker2019,
       author = {{Stolker}, T. and {Bonse}, M.~J. and {Quanz}, S.~P. and {Amara}, A. and {Cugno}, G. and {Bohn}, A.~J. and {Boehle}, A.},
        title = "{PynPoint: a modular pipeline architecture for processing and analysis of high-contrast imaging data}",
      journal = {\aap},
         year = 2019,
        month = jan,
       volume = {621},
          eid = {A59},
        pages = {A59},
          doi = {10.1051/0004-6361/201834136},
archivePrefix = {arXiv},
       eprint = {1811.03336},
 primaryClass = {astro-ph.EP},
       adsurl = {https://ui.adsabs.harvard.edu/abs/2019A&A...621A..59S}
}

@ARTICLE{Christiaens2021,
       author = {{Christiaens}, V. and {Ubeira-Gabellini}, M. -G. and {C{\'a}novas}, H. and {Delorme}, P. and {Pairet}, B. and {Absil}, O. and {Casassus}, S. and {Girard}, J.~H. and {Zurlo}, A. and {Aoyama}, Y. and {Marleau}, G. -D. and {Spina}, L. and {van der Marel}, N. and {Cieza}, L. and {Lodato}, G. and {P{\'e}rez}, S. and {Pinte}, C. and {Price}, D.~J. and {Reggiani}, M.},
        title = "{A faint companion around CrA-9: protoplanet or obscured binary?}",
      journal = {\mnras},
         year = 2021,
        month = apr,
       volume = {502},
       number = {4},
        pages = {6117-6139},
          doi = {10.1093/mnras/stab480},
archivePrefix = {arXiv},
       eprint = {2102.10288},
 primaryClass = {astro-ph.EP},
       adsurl = {https://ui.adsabs.harvard.edu/abs/2021MNRAS.502.6117C}
}

@ARTICLE{christiaens2023b,
       author = {{Christiaens}, Valentin and {Hammond}, Iain and {Juillard}, Sandrine and {Kokoulina}, Elena and {Balsalobre-Ruza}, Olga},
        title = "{VCAL-SPHERE: Hybrid pipeline for reduction of VLT/SPHERE data}",
 howpublished = {Astrophysics Source Code Library, record ascl:2311.002},
         year = 2023,
        month = nov,
          eid = {ascl:2311.002},
archivePrefix = {ascl},
       eprint = {2311.002},
       adsurl = {https://ui.adsabs.harvard.edu/abs/2023ascl.soft11002C}
}

@ARTICLE{Aoyama2018,
       author = {{Aoyama}, Yuhiko and {Ikoma}, Masahiro and {Tanigawa}, Takayuki},
        title = "{Theoretical Model of Hydrogen Line Emission from Accreting Gas Giants}",
      journal = {\apj},
         year = 2018,
        month = oct,
       volume = {866},
       number = {2},
          eid = {84},
        pages = {84},
          doi = {10.3847/1538-4357/aadc11},
archivePrefix = {arXiv},
       eprint = {1808.06776},
 primaryClass = {astro-ph.EP},
       adsurl = {https://ui.adsabs.harvard.edu/abs/2018ApJ...866...84A}
}

@ARTICLE{vanCap2025,
       author = {{van Capelleveen}, Richelle F. and {Ginski}, Christian and {Kenworthy}, Matthew A. and {Byrne}, Jake and {Lawlor}, Chloe and {McLachlan}, Dan and {Mamajek}, Eric E. and {Stolker}, Tomas and {Benisty}, Myriam and {Bohn}, Alexander J. and {Close}, Laird M. and {Dominik}, Carsten and {Haffert}, Sebastiaan and {Landman}, Rico and {Ma}, Jie and {Snellen}, Ignas and {Tazaki}, Ryo and {van der Marel}, Nienke and {Welzel}, Lukas and {Zhang}, Yapeng},
        title = "{WIde Separation Planets In Time (WISPIT): A Gap-clearing Planet in a Multi-ringed Disk around the Young Solar-type Star WISPIT 2}",
      journal = {\apjl},
         year = 2025,
        month = sep,
       volume = {990},
       number = {1},
          eid = {L8},
        pages = {L8},
          doi = {10.3847/2041-8213/adf721},
archivePrefix = {arXiv},
       eprint = {2508.19053},
 primaryClass = {astro-ph.EP},
       adsurl = {https://ui.adsabs.harvard.edu/abs/2025ApJ...990L...8V}
}

@ARTICLE{Stapper2022,
       author = {{Stapper}, L.~M. and {Ginski}, C.},
        title = "{Iterative angular differential imaging (IADI): An exploration of recovering disk structures in scattered light with an iterative ADI approach}",
      journal = {\aap},
         year = 2022,
        month = dec,
       volume = {668},
          eid = {A50},
        pages = {A50},
          doi = {10.1051/0004-6361/202142820},
archivePrefix = {arXiv},
       eprint = {2210.02454},
 primaryClass = {astro-ph.IM},
       adsurl = {https://ui.adsabs.harvard.edu/abs/2022A&A...668A..50S}
}

@ARTICLE{Garufi2018,
       author = {{Garufi}, A. and {Benisty}, M. and {Pinilla}, P. and {Tazzari}, M. and {Dominik}, C. and {Ginski}, C. and {Henning}, Th. and {Kral}, Q. and {Langlois}, M. and {M{\'e}nard}, F. and {Stolker}, T. and {Szulagyi}, J. and {Villenave}, M. and {van der Plas}, G.},
        title = "{Evolution of protoplanetary disks from their taxonomy in scattered light: spirals, rings, cavities, and shadows}",
      journal = {\aap},
         year = 2018,
        month = dec,
       volume = {620},
          eid = {A94},
        pages = {A94},
          doi = {10.1051/0004-6361/201833872},
archivePrefix = {arXiv},
       eprint = {1810.04564},
 primaryClass = {astro-ph.SR},
       adsurl = {https://ui.adsabs.harvard.edu/abs/2018A&A...620A..94G}
}

@ARTICLE{Dong2018,
       author = {{Dong}, Ruobing and {Najita}, Joan R. and {Brittain}, Sean},
        title = "{Spiral Arms in Disks: Planets or Gravitational Instability?}",
      journal = {\apj},
         year = 2018,
        month = aug,
       volume = {862},
       number = {2},
          eid = {103},
        pages = {103},
          doi = {10.3847/1538-4357/aaccfc},
archivePrefix = {arXiv},
       eprint = {1806.05183},
 primaryClass = {astro-ph.SR},
       adsurl = {https://ui.adsabs.harvard.edu/abs/2018ApJ...862..103D}
}

@ARTICLE{Xie2022,
       author = {{Xie}, Chen and {Choquet}, Elodie and {Vigan}, Arthur and {Cantalloube}, Faustine and {Benisty}, Myriam and {Boccaletti}, Anthony and {Bonnefoy}, Mickael and {Desgrange}, Celia and {Garufi}, Antonio and {Girard}, Julien and {Hagelberg}, Janis and {Janson}, Markus and {Kenworthy}, Matthew and {Lagrange}, Anne-Marie and {Langlois}, Maud and {Menard}, Fran{\c{c}}ois and {Zurlo}, Alice},
        title = "{Reference-star differential imaging on SPHERE/IRDIS}",
      journal = {\aap},
         year = 2022,
        month = oct,
       volume = {666},
          eid = {A32},
        pages = {A32},
          doi = {10.1051/0004-6361/202243379},
archivePrefix = {arXiv},
       eprint = {2208.07915},
 primaryClass = {astro-ph.EP},
       adsurl = {https://ui.adsabs.harvard.edu/abs/2022A&A...666A..32X}
}

@INPROCEEDINGS{Delorme2017,
       author = {{Delorme}, P. and {Meunier}, N. and {Albert}, D. and {Lagadec}, E. and {Le Coroller}, H. and {Galicher}, R. and {Mouillet}, D. and {Boccaletti}, A. and {Mesa}, D. and {Meunier}, J. -C. and {Beuzit}, J. -L. and {Lagrange}, A. -M. and {Chauvin}, G. and {Sapone}, A. and {Langlois}, M. and {Maire}, A. -L. and {Montarg{\`e}s}, M. and {Gratton}, R. and {Vigan}, A. and {Surace}, C.},
        title = "{The SPHERE Data Center: a reference for high contrast imaging processing}",
    booktitle = {SF2A-2017: Proceedings of the Annual meeting of the French Society of Astronomy and Astrophysics},
         year = 2017,
       editor = {{Reyl{\'e}}, C. and {Di Matteo}, P. and {Herpin}, F. and {Lagadec}, E. and {Lan{\c{c}}on}, A. and {Meliani}, Z. and {Royer}, F.},
        month = dec,
        pages = {Di},
          doi = {10.48550/arXiv.1712.06948},
archivePrefix = {arXiv},
       eprint = {1712.06948},
 primaryClass = {astro-ph.IM},
       adsurl = {https://ui.adsabs.harvard.edu/abs/2017sf2a.conf..347D}
}

@ARTICLE{Montesinos2016,
       author = {{Montesinos}, Mat{\'\i}as and {Perez}, Sebastian and {Casassus}, Simon and {Marino}, Sebastian and {Cuadra}, Jorge and {Christiaens}, Valentin},
        title = "{Spiral Waves Triggered by Shadows in Transition Disks}",
      journal = {\apjl},
         year = 2016,
        month = may,
       volume = {823},
       number = {1},
          eid = {L8},
        pages = {L8},
          doi = {10.3847/2041-8205/823/1/L8},
archivePrefix = {arXiv},
       eprint = {1601.07912},
 primaryClass = {astro-ph.EP},
       adsurl = {https://ui.adsabs.harvard.edu/abs/2016ApJ...823L...8M}
}

@ARTICLE{Montesinos2018,
       author = {{Montesinos}, Mat{\'\i}as and {Cuello}, Nicol{\'a}s},
        title = "{Planetary-like spirals caused by moving shadows in transition discs}",
      journal = {\mnras},
         year = 2018,
        month = mar,
       volume = {475},
       number = {1},
        pages = {L35-L39},
          doi = {10.1093/mnrasl/sly001},
archivePrefix = {arXiv},
       eprint = {1712.09157},
 primaryClass = {astro-ph.EP},
       adsurl = {https://ui.adsabs.harvard.edu/abs/2018MNRAS.475L..35M}
}

@ARTICLE{vdM2016,
       author = {{van der Marel}, N. and {Cazzoletti}, P. and {Pinilla}, P. and {Garufi}, A.},
        title = "{Vortices and Spirals in the HD135344B Transition Disk}",
      journal = {\apj},
         year = 2016,
        month = dec,
       volume = {832},
       number = {2},
          eid = {178},
        pages = {178},
          doi = {10.3847/0004-637X/832/2/178},
archivePrefix = {arXiv},
       eprint = {1607.05775},
 primaryClass = {astro-ph.EP},
       adsurl = {https://ui.adsabs.harvard.edu/abs/2016ApJ...832..178V}
}

@ARTICLE{Cazzoletti2018,
       author = {{Cazzoletti}, P. and {van Dishoeck}, E.~F. and {Pinilla}, P. and {Tazzari}, M. and {Facchini}, S. and {van der Marel}, N. and {Benisty}, M. and {Garufi}, A. and {P{\'e}rez}, L.~M.},
        title = "{Evidence for a massive dust-trapping vortex connected to spirals. Multi-wavelength analysis of the HD 135344B protoplanetary disk}",
      journal = {\aap},
         year = 2018,
        month = nov,
       volume = {619},
          eid = {A161},
        pages = {A161},
          doi = {10.1051/0004-6361/201834006},
archivePrefix = {arXiv},
       eprint = {1809.04160},
 primaryClass = {astro-ph.EP},
       adsurl = {https://ui.adsabs.harvard.edu/abs/2018A&A...619A.161C}
}

@ARTICLE{Maire2016,
       author = {{Maire}, Anne-Lise and {Langlois}, Maud and {Dohlen}, Kjetil and {Lagrange}, Anne-Marie and {Gratton}, Raffaele and {Chauvin}, Ga{\"e}l. and {Desidera}, Silvano and {Girard}, Julien H. and {Milli}, Julien and {Vigan}, Arthur and {Zins}, Gerard and {Delorme}, Philippe and {Beuzit}, Jean-Luc and {Claudi}, Riccardo U. and {Feldt}, Markus and {Mouillet}, David and {Puget}, Pascal and {Turatto}, Massimo and {Wildi}, Fran{\c{c}}ois},
        title = "{SPHERE IRDIS and IFS astrometric strategy and calibration}",
    booktitle = {Ground-based and Airborne Instrumentation for Astronomy VI},
         year = 2016,
       editor = {{Evans}, Christopher J. and {Simard}, Luc and {Takami}, Hideki},
       series = {Society of Photo-Optical Instrumentation Engineers (SPIE) Conference Series},
       volume = {9908},
        month = aug,
          eid = {990834},
        pages = {990834},
          doi = {10.1117/12.2233013},
archivePrefix = {arXiv},
       eprint = {1609.06681},
 primaryClass = {astro-ph.IM},
       adsurl = {https://ui.adsabs.harvard.edu/abs/2016SPIE.9908E..34M}
}

@manual{ESO_ERIS_Manual,
    author = {ESO},
  title        = {ERIS User Manual},
  organization = {European Southern Observatory},
  year         = {2024},
  url          = {https://www.eso.org/sci/facilities/paranal/instruments/eris/doc.html},
    note         = {Version 116.0}
}

@ARTICLE{Launhardt2020,
       author = {{Launhardt}, R. and {Henning}, Th. and {Quirrenbach}, A. and {S{\'e}gransan}, D. and {Avenhaus}, H. and {van Boekel}, R. and {Brems}, S.~S. and {Cheetham}, A.~C. and {Cugno}, G. and {Girard}, J. and {Godoy}, N. and {Kennedy}, G.~M. and {Maire}, A.-L. and {Metchev}, S. and {M{\"u}ller}, A. and {Musso Barcucci}, A. and {Olofsson}, J. and {Pepe}, F. and {Quanz}, S.~P. and {Queloz}, D. and {Reffert}, S. and {Rickman}, E.~L. and {Ruh}, H.~L. and {Samland}, M.},
        title = "{ISPY-NACO Imaging Survey for Planets around Young stars. Survey description and results from the first 2.5 years of observations}",
      journal = {\aap},
         year = 2020,
        month = mar,
       volume = {635},
          eid = {A162},
        pages = {A162},
          doi = {10.1051/0004-6361/201937000},
archivePrefix = {arXiv},
       eprint = {2002.01807},
 primaryClass = {astro-ph.EP},
       adsurl = {https://ui.adsabs.harvard.edu/abs/2020A&A...635A.162L}
}

@ARTICLE{Godoy2022,
       author = {{Godoy}, N. and {Olofsson}, J. and {Bayo}, A. and {Cheetham}, A.~C. and {Launhardt}, R. and {Chauvin}, G. and {Kennedy}, G.~M. and {Brems}, S.~S. and {Cugno}, G. and {Girard}, J.~H. and {Henning}, Th. and {M{\"u}ller}, A. and {Barcucci}, A. Musso and {Pepe}, F. and {Quanz}, S.~P. and {Quirrenbach}, A. and {Reffert}, S. and {Rickman}, E.~L. and {Samland}, M. and {S{\'e}gransan}, D. and {Stolker}, T.},
        title = "{ISPY - NaCo Imaging Survey for Planets around Young stars. CenteR: The impact of centering and frame selection}",
      journal = {\aap},
         year = 2022,
        month = jul,
       volume = {663},
          eid = {A53},
        pages = {A53},
          doi = {10.1051/0004-6361/202040024},
archivePrefix = {arXiv},
       eprint = {2111.14888},
 primaryClass = {astro-ph.IM},
       adsurl = {https://ui.adsabs.harvard.edu/abs/2022A&A...663A..53G}
}

@INPROCEEDINGS{Orban2024,
       author = {{Orban de Xivry}, G. and {Absil}, O. and {De Rosa}, R.~J. and {Bonse}, M.~J. and {Dannert}, F. and {Hayoz}, J. and {Grani}, P. and {Puglisi}, A. and {Baruffolo}, A. and {Salasnich}, B. and {Davies}, R. and {Glauser}, A.~M. and {Huby}, E. and {Kenworthy}, M. and {Quanz}, S.~P. and {Taylor}, W. and {Zins}, G.},
        title = "{The VLT/ERIS vortex coronagraph: design, pointing control, and on-sky performance}",
    booktitle = {Adaptive Optics Systems IX},
         year = 2024,
       editor = {{Jackson}, Kathryn J. and {Schmidt}, Dirk and {Vernet}, Elise},
       series = {Society of Photo-Optical Instrumentation Engineers (SPIE) Conference Series},
       volume = {13097},
        month = aug,
          eid = {1309715},
        pages = {1309715},
          doi = {10.1117/12.3018265},
archivePrefix = {arXiv},
       eprint = {2407.14406},
 primaryClass = {astro-ph.IM},
       adsurl = {https://ui.adsabs.harvard.edu/abs/2024SPIE13097E..15O}
}

@ARTICLE{Maire2021,
       author = {{Maire}, Anne-Lise and {Langlois}, Maud and {Delorme}, Philippe and {Chauvin}, Ga{\"e}l. and {Gratton}, Raffaele and {Vigan}, Arthur and {Girard}, Julien H. and {Wahhaj}, Zahed and {Pott}, J{\"o}rg-Uwe and {Burtscher}, Leonard and {Boccaletti}, Anthony and {Carlotti}, Alexis and {Henning}, Thomas and {Kenworthy}, Matthew A. and {Kervella}, Pierre and {Rickman}, Emily L. and {Schmidt}, Tobias O.~B.},
        title = "{Lessons learned from SPHERE for the astrometric strategy of the next generation of exoplanet imaging instruments}",
      journal = {Journal of Astronomical Telescopes, Instruments, and Systems},
         year = 2021,
        month = jul,
       volume = {7},
          eid = {035004},
        pages = {035004},
          doi = {10.1117/1.JATIS.7.3.035004},
archivePrefix = {arXiv},
       eprint = {2107.14341},
 primaryClass = {astro-ph.IM},
       adsurl = {https://ui.adsabs.harvard.edu/abs/2021JATIS...7c5004M}
}

\clearpage

\begin{appendix}

\section{JWST candidate}\label{app:JWST}
\cite{cugno2024} presented a candidate protoplanet at a separation of 2\farcs2, detected with JWST in the 410M band only. Our reprocessing of the data, after a fine recentering of the images, recovers the candidate detection, with an SNR of 2.4 only. This candidate is unlikely to be the object driving the spirals as it does not match the orbital velocity measurements presented in \cite{xie2024} and in this work, though that is not enough to rule out this hypothetical companion.

\begin{figure}[h!]
    \centering
    \includegraphics[width=0.40\textwidth]{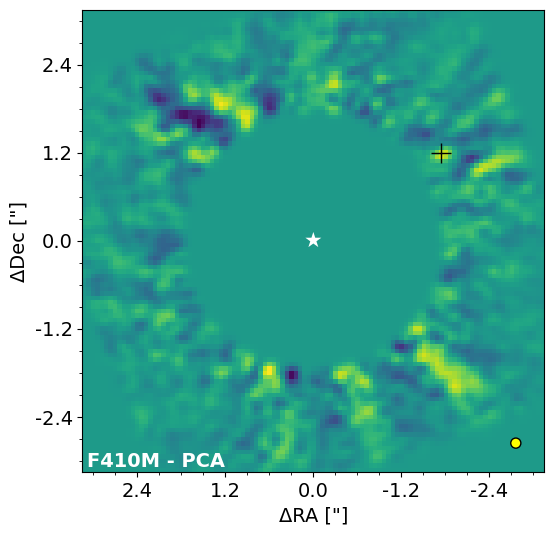}
    \caption{PCA image in the F410M filter captured with JWST-NIRCam. The candidate found by \citet{cugno2024} is marked by the black cross.}
    \label{fig:F410M}
\end{figure}

\FloatBarrier

\section{Image Gallery}
\begin{figure}[h!]
    \centering
    \includegraphics[width=0.49\textwidth]{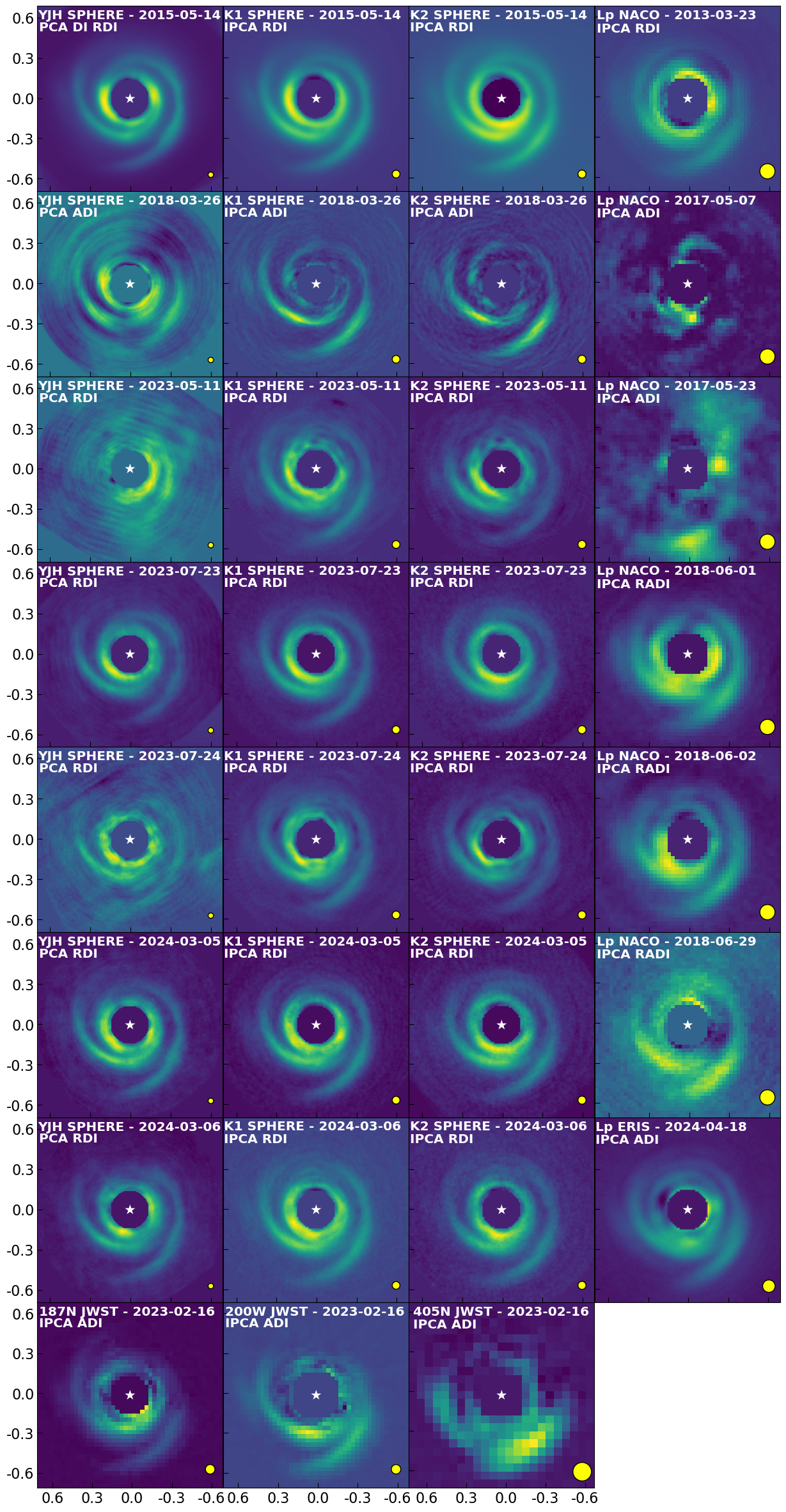}
    \caption{Gallery of the best disk images obtained for all datasets at all wavelengths bands available. \new{The white star marks the center of the images and the yellow circles illustrate the resolution of each image.}}
    \label{fig:FullImageGallery}
\end{figure}

All the datasets in our possession were systematically processed with algorithms that try to minimize as much as possible the biases related to ADI and preserve as well as possible the morphology of the disk. Figure~\ref{fig:FullImageGallery} shows the full gallery of images. Strategies to obtain the highest-fidelity images of the disk included using RDI when possible, combined with IPCA. Data imputation was also used, for the 2015 SPHERE dataset only.

We notice that the datasets with the smallest field rotation produce the worst images of the disk, especially if there is no reference star to enable an RDI strategy, in which case the disk is significantly self-subtracted and unrecoverable with our algorithms. This is the case for the two NACO observations captured in 2017. Even when we only use RDI and no ADI, we find that a sufficient field rotation is still beneficial, as illustrated by the difference in quality between the SPHERE 2024-03-06 (58$^{\circ}$ rotation) and SPHERE 2023-07-24 (12$^{\circ}$ rotation), despite similar seeing and observing time.

\FloatBarrier

\section{True North Correction} \label{app:TN}
As the True North (TN) corrections are crucial for an accurate measurement of the spirals' dynamics, Table~\ref{tab:TNCorrections} contains the TN values we used for each instrument.

\begin{table}[h!] 
\begin{center}
\caption{True North corrections applied.} 
\label{tab:TNCorrections}
\begin{tabular}{lcc}
\hline
\hline
Instrument & TN correction (\degr) & Reference \\ \hline
NACO & $0.572 \pm 0.178$ & \cite{Launhardt2020} \\
SPHERE & $-1.75 \pm 0.08$ & \cite{Maire2016}\\
ERIS & $4.670 \pm 0.016$ & \cite{ESO_ERIS_Manual}\\

\hline
\end{tabular}
\end{center}
\end{table}

\FloatBarrier

\begin{figure}[h!]
    \centering
    \includegraphics[width=0.45\textwidth]{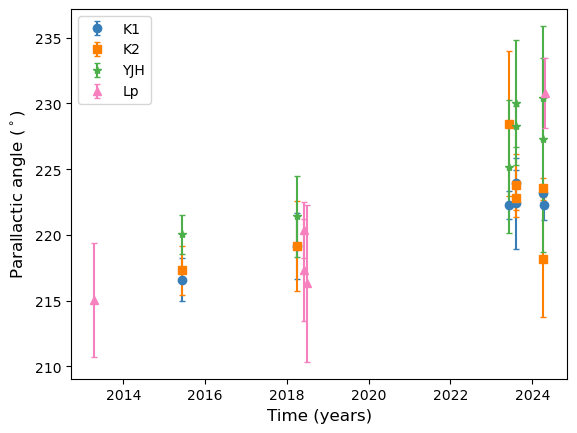}
    \caption{Position of the twist found in S1 as a function of time.}
    \label{fig:TwistPosition}
\end{figure}

\newbis{
\section{Twist position}
We measured the position of the twist in every dataset and wavelength bands, except for the two NACO observations in 2017 that do not detect the disk well enough. Fig. \ref{fig:TwistPosition} shows its position, with the associated uncertainties, for all epochs and wavelengths bands, illustrating its motion over time.}

\FloatBarrier

\begin{figure}[h!]
    \centering
    \includegraphics[width=0.4\textwidth]{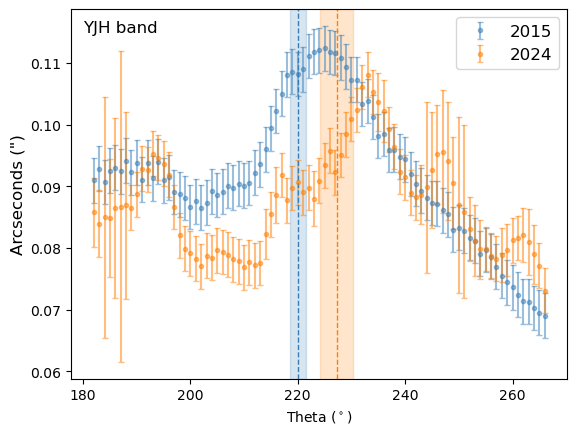}
    \includegraphics[width=0.4\textwidth]{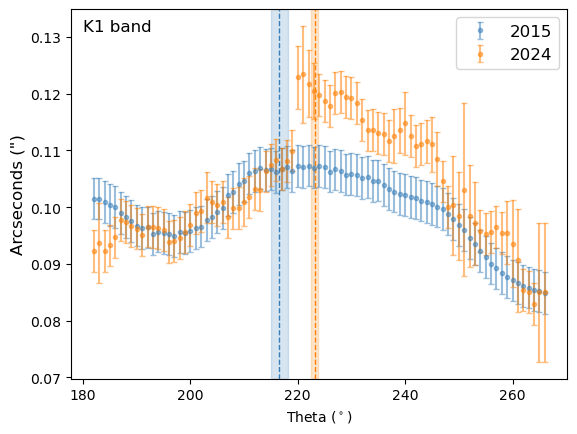}
    \includegraphics[width=0.4\textwidth]{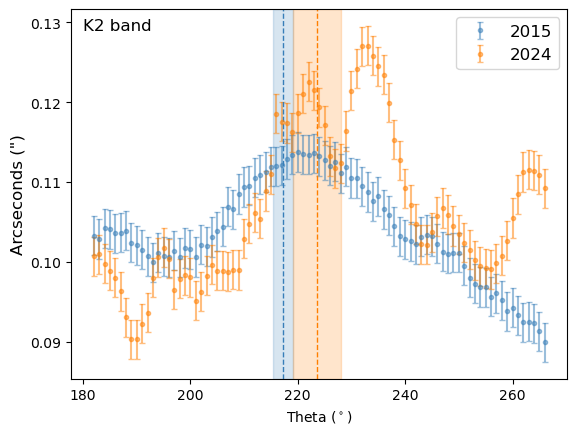}
    \includegraphics[width=0.4\textwidth]{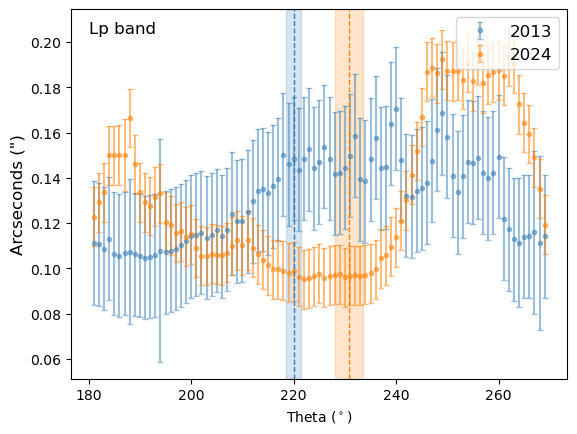}
    \caption{Measured FWHM of the S1 arm at all available wavelengths. The vertical lines along with the shaded areas on either side represent the twist position and its associated uncertainties.}
    \label{fig:BlobPosition}
\end{figure}

\begin{figure*}[h!]
    \centering
    \includegraphics[width=0.89\textwidth]{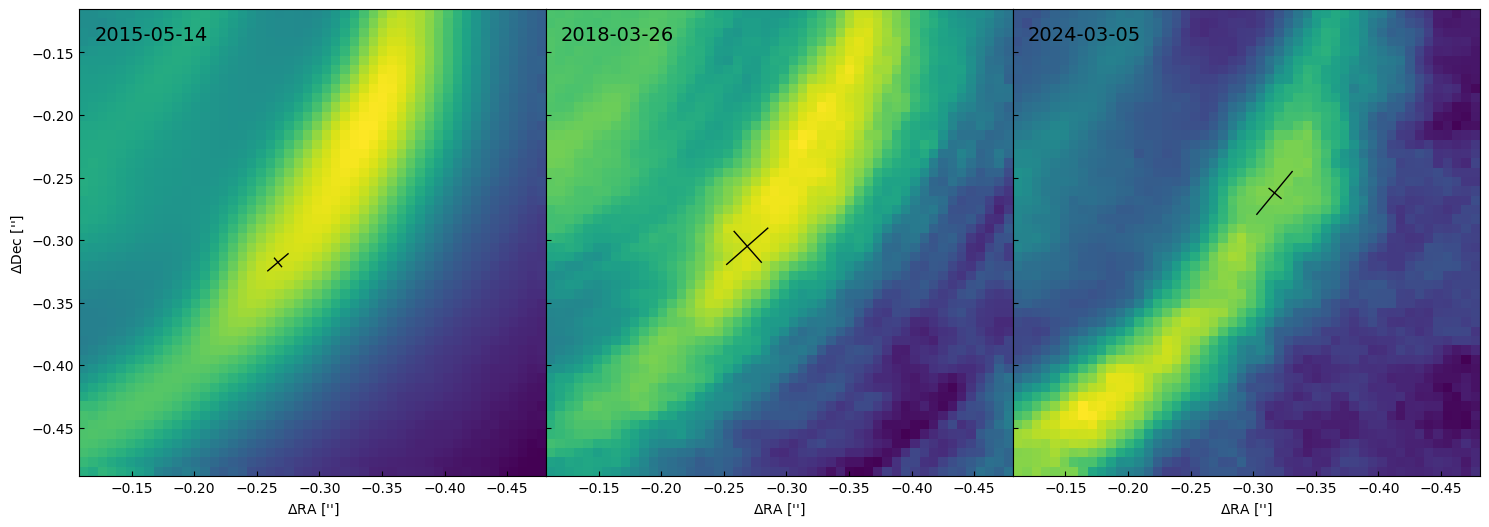}
    \caption{Close-up of the blob in YJH band in three different epochs. The location of the twist and its associated uncertainties are marked by the black cross, visualizing the apparent connection between the twist and the wider section of S1.}
    \label{fig:TwistVisualization}
\end{figure*}

\newthree{
\section{S1 blob motion}
We measured the FWHM of the S1 arm in every dataset and found a correlation between the twist position and the wider section of the S1 arm. However, we were not able to show conclusively that this blob is moving along with the twist, as shown in Fig. \ref{fig:BlobPosition}, though there seems to be some hint of motion in the same direction expected as the twist. This is visualized in Fig. \ref{fig:TwistVisualization}. The uncertainties on our measurements and the variations from one dataset to another, probably due to the recovery of the disk signal not being perfect with IPCA, or due to variable shadowing effects on the disk, prevent a rigorous measurement of the blob motion. Additional high-quality observations and observing time comparable to the 2015 dataset would be needed to obtain more robust measurements. We do not reliably detect this blob in Lp band, while it is best seen in YJH band, which we attribute to the angular resolution.}

\end{appendix}

\end{document}